\documentstyle[11pt,aaspp4]{article}

\def\degr{\hbox{$^\circ$}}
\def\Sec{${}^{\prime\prime}$\llap{.}}

\slugcomment{To appear in Ap J.}

\lefthead{Mendes de Oliveira et al.}
\righthead{Ionized gas content in Hickson
Compact Group 16}

\begin{document}

\title{Extended ionized gas emission and kinematics of the compact group
galaxies in HCG 16: Signatures of mergers}

\author{C. Mendes de Oliveira\altaffilmark{1}}
\affil{Instituto Astron\^omico e Geof\'{\i}sico (IAG),
       Av Miguel Stefano 4200,  04301-904, S\~ao Paulo, Brazil}

\author{H.~Plana\altaffilmark{2} and P.~Amram}
\affil{Observatoire de Marseille,
       2 Place Le Verrier, 13248 Marseille, Cedex 04, France}

\author{M.~Bolte}
\affil{ Lick Observatory,
       Board of Studies in Astronomy and Astrophysics, University of
       California, Santa Cruz, California 95064}

\and

\author{J.~Boulesteix}
\affil{Observatoire de Marseille,
       2 Place Le Verrier, 13248 Marseille, Cedex 04, France}
 

\altaffiltext{1}{present address: 
Universit\"ats-Sternwarte, Ludwig-Maximilians-Universit\"at,
       Scheinersstrasse 1, 81679 Muenchen, Germany}
\altaffiltext{2}{present address: Instituto Astron\^omico e
Geof\'{\i}sico (IAG),
Av Miguel Stefano 4200, CEP: 04301-904, S\~ao Paulo, Brazil}


\begin{abstract}

   We report on kinematic observations of $H\alpha $ emission from four
late-type galaxies of the Hickson Compact Group 16 (H16 a,b,c and d)
obtained with a scanning Fabry-Perot interferometer and samplings of 16
$km$ $s^{-1}$ and 1$\arcsec$.  The velocity fields show kinematic
peculiarities for three of the four galaxies:  H16b, c and d.
Misalignments between the kinematic and photometric axes of gas and stellar
components (H16b,c,d),  double gas systems (H16c) and severe warping of
the kinematic major axis (H16b and c) were some of the peculiarities
detected.

    We conclude that major merger events have taken place in at least
two of the galaxies of the group, H16c and d, based on their
significant kinematic peculiarities, their double nuclei and high
infrared luminosities.  Their H$\alpha$ gas content is strongly
spatially concentrated -- H16d contains a peculiar bar-like structure
confined to the inner $\sim$ 1 h$^{-1}$ kpc region.  These observations
are in agreement with predictions of simulations, namely that the gas
flows towards the galaxy nucleus during mergers, forms bars and fuels
the central activity.  Galaxy H16b, an Sb galaxy, also presents some of
the kinematic evidences for past accretion events.  Its gas content,
however, is very sparse, limiting our ability to find other kinematic
merging indicators, if they are present.  We find that the merger
remnants in the compact group HCG 16 have significantly
{\it smoother} optical profiles than isolated mergers, i.e., they show 
an amorphous morphology and no signs of tidal tails. Tidal arms and tails 
formed during the mergers may have been stripped by the
group potential (\cite{bar92}) or alternatively
they may have never been formed.

   The velocity field of the galaxy H16a shows grand-design isovelocity
lines with no signs of disturbances inside a radius of $\sim$
R$_{25}$.  This result is contrary to expectations given that the
galaxy has a high infrared luminosity, central activity, tidal
tails at large radii and it is embedded in a common group envelope
observed in HI and x-rays. The normality of the velocity field suggests
that this galaxy may be a fairly recent acquisition of the group.
  
  Our observations suggest that HCG 16 may be a young compact group in
formation through the merging of close-by objects in a dense
environment.
\end{abstract}




%

\section{Introduction}

   Nearby compact groups may provide one of the best laboratories for
studying the effects of on-going collisions on the structure and
dynamics of galaxies.  One of the most dense compact groups known is
HGC 16 (H16), or Arp 318, originally catalogued by \cite{hic82} as an
association of four late-type galaxies at V $\sim$ 4000 $km s^{-1}$
with a small velocity dispersion of 123 $km s^{-1}$ and a median
galaxy-galaxy projected separation of 44 h$^{-1}$ kpc (\cite{hic92},
$h$ is the dimensionless Hubble constant $H_o$/100 km sec$^{-1}$
Mpc$^{-1}$).  \cite{rib96} found three other bright galaxies near the
group, in the same redshift range, within a median radius of 0.197
h$^{-1}$ Mpc. All seven galaxies were found to have nuclear emission
lines, making H16 the highest concentration of starburst and active
nuclei galaxies in the nearby universe.

    ROSAT observations of the group (\cite{pon96}) revealed an
intragroup medium with a total extent of $\sim$ 400 h$^{-1}$ kpc, a
temperature of 0.3 KeV and density of N$_H$ = 2.05 $\times$ 10$^{20}$
cm$^{-1}$.  Such density and temperature are typical of compact groups
of galaxies, which were shown to fit well on the extension of the
correlations between the X-ray luminosity and both the gas temperature
and velocity dispersion of clusters of galaxies (although with a
steeper slope than for clusters (\cite{pon96}).  The presence of
diffuse X-ray emission around H16 is a strong evidence that this is a
genuine group.

   Compact groups that are true high-density systems, as H16 seems to
be, provide excellent environments in which to study the properties of
merger galaxies.  Such a study for a compact group galaxy is important
not only to investigate the process of galaxy formation through mergers
but it may also provide important information on the dynamical
evolutionary stage of the compact group.

  In this paper we investigate the dynamical state of the Hickson
compact group 16 through the study of the detailed 2-D kinematics of
four of its member galaxies:  H16a (NGC 835, Mrk 1022), H16b (NGC 833),
H16c (NGC 838, Mrk 1021) and H16d (NGC 839).  The four galaxies are
optically luminous systems, with M$_B$ ranging from --20.3 $+$ 5 log $h$
to --19.5 $+$ 5 log $h$. H16a, c and d are strong radio and infrared
sources (\cite{all96}, \cite{men95}). \cite{bos96} report
on CO observations for three of the four galaxies and calculate large
masses of molecular gas.  Analysis of their central spectra shows that
H16a and d are LINERs, H16b is a Seyfert galaxy and H16c 
has a starburst nucleus (\cite{coz98}). Rubin, Hunter \& Ford
(1991, RHF91) obtained rotation curves (RCs) for the four galaxies and
found they have ``abnormal'' shapes, a common feature among strongly
interacting and merging galaxies.

  H$\alpha$ emission-line observations of H16 a,b,c,d obtained with a
scanning Fabry-Perot are used here to derive the spatial distribution
and kinematics of their warm gas content.  The warm gas is a good
tracer of the total gas mass (that is gas in all phases) in a galaxy
(\cite{sch93}) and it should also be a good tracer of the
potential of the galaxy.  The details of the kinematics of the
emission-line velocity field can be used as diagnostics of the gaseous
merger dynamical state.

   The organization of this article is as follows:  Section 2 gives details
of the observations and data reduction.  In Section 3, we present the
results for the photometry, the internal kinematics and the various
mass determinations of the galaxies.  Section 4 contains the
discussion and our conclusions.

\section{Observations}

\subsection{Photometric data}

Images of H16a,b,c,d, in R and I were available for this study.  The
CCD R-band images were taken as part of a large study of the systematic
properties of galaxies in compact groups.  Details of these data are
described in \cite{hic89}. The images have kindly
been made available by P. Hickson.  The I-band image was a 600s--frame
(15' on a side, 0.44 arcsec pixels) taken with the 1.5m telescope at
the Cerro Tololo Interamerican Observatory in October 1997. The mean
seeing values for the R and I images were 1\Sec5 and 1\Sec2
respectively.

   Luminosity, ellipticity and position angle (PA) profiles were
determined for H16b, c and d, using the ISOPHOTE package in STSDAS
(\cite{jed87}).  H16a is not well described by concentric
elliptical isophotes and therefore no fit of this galaxy was
attempted.

\subsection{Fabry-Perot data}

  Observations were carried out with the Fabry-Perot instrument CIGALE
mounted on the ESO 3.6m telescope in August 1995. CIGALE is composed of
a focal reducer (bringing the original f/8 focal ratio of the
Cassegrain focus to f/2), a scanning Fabry-Perot, a narrow band
interference filter and an Image Photon Counting System 
detector (IPCS).  The IPCS, with a time sampling of 1/50 s and zero readout
noise makes it possible to scan the interferometer rapidly (typically 5
s per channel) avoiding sky transparency, airmass and seeing variation
problems during the exposures.

  Table 1 contains the journal of the observations.  The exposure times
were two hours for each of two pointings, one on H16ab and one on
H16cd. The two observed fields are represented with boxes in Fig. 1.
Table 2 presents general information on the four observed galaxies.

\placetable{tbl-1}

\placetable{tbl-2}
  
  Reduction of the data cubes were performed using the CIGALE/ADHOC
software (\cite{bou93}).  The data reduction procedure has been
extensively described in \cite{amr96} and references
therein.

  Wavelength calibrations were obtained by scanning the narrow Ne 6599
\AA\ line under the same conditions as the observations.  The relative
velocities with respect to the systemic velocity are very accurate,
with an error of a fraction of a channel width (${\rm <3 \, km
s^{-1}}$) over the whole field.

    The signal measured along the scanning sequence was separated
into two parts: (1) an almost constant level produced by the
continuum light in a 24 \AA ~passband around H$\alpha $ (continuum
map), and (2) a varying part produced by the H$\alpha $ line
(monochromatic map).  The continuum level was taken to be the mean of
the three faintest channels, to avoid channel noise effects.  The
monochromatic map was obtained by integrating the monochromatic profile
in each pixel.  The velocity sampling was 16 $km$ $s^{-1}$.  The
monochromatic maps had one-pixel resolution in the center of the
galaxies. Spectral profiles were binned in the outer parts (to 5x5
pixels) in order to increase the signal-to-noise ratio.   OH
night sky lines passing through the filters were subtracted by
determining the level of emission away from the galaxies
(\cite{lav87}).

  A rough flux calibration of the monochromatic images  was attempted
by adjusting the flux levels to those of the calibrated image of the
Cartwheel galaxy, obtained in the same run (see details of how the
Cartwheel galaxy image was calibrated in \cite{amr98}). H$\alpha$
profiles for the H16 galaxies were measured to a minimum flux density
between 0.1 $\times$ 10$^{-17}$ erg s$^{-1}$cm$^{-2}$ arcsec$^{-2}$ and
0.7 $\times$ 10$^{-17}$ erg s$^{-1}$cm$^{-2}$ arcsec$^{-2}$
(corresponding to a S/N between three and eight).


%

\section{Results}

\subsection{Morphology from the R, I and continuum images}

Fig. 1 is the I-band image of the group with contours obtained from the 
continuum images superimposed.

\placefigure{fig01}

\subsubsection{H16ab}
  
   H16a and H16b are on the top right of Fig. 1 (H16a is the galaxy
further to the east).  They have been morphologically classified as
early-type spiral galaxies (SBab and Sab respectively) by \cite{hic93}
from inspection of CCD images.  This classification is in
general agreement with morphological classifications given by RHF91 and
the RC3.  The morphologies of H16a and H16b have been described in
detail by RHF91.  We confirm the central dust and the low-surface
brightness tidal tails to the east (and a shorter one to the south) of
H16a extending over 20 h$^{-1}$ kpc. 
\cite{hun96} detected three dwarf galaxy candidates on
the eastern tail of H16a (discussed in Section 4.3).
For H16b, the subtraction of a
model galaxy with concentric elliptical isophotes uncovered an 
asymmetric cone-like structure to the west of the galaxy.

\subsubsection{H16c}  

  H16c and H16d, on the top and lower left of Fig. 1, have been
classified as Irregular galaxies by \cite{hic93} and as S0 pec by
\cite{dev91} (the RC3 catalog).  The amorphous shapes of these galaxies
make it difficult to visually determine if they are either bulge or
disk dominated systems.  R-band images of the central parts of these
two galaxies are shown in Fig. 2.

\placefigure{fig02}

H16c has two bright centers $\sim$ 2$\arcsec$ apart
(see Fig. 2a).  The resolution of the
Fabry-Perot image (Fig. 9a) is too low to separate the two  
nuclei.  Neither of the two bright nuclei coincide with the
center of the outer  (outside of 10$\arcsec$) isophotes of the galaxy.  A
bright extension of $\sim$ 5$\arcsec$ directly west from both centers
is also detected (see Fig. 2a).  This extension is seen in the R,
I, continuum and monochromatic images.

   Fig. 3 presents the surface brightness profiles of H16c obtained by
fitting the galaxy with concentric elliptical isophotes. Profiles are
obtained from the emission-line image and continuum-image in order to
check if reasonable fits can be found to exponential or de Vaucouleurs
r$^{1/4}$ profiles.  The data are plotted in linear scale (radius in
arcsec) for the two curves to the left of the diagram and in radius to
the power $1/4$, for the two curves to the right of the diagram.  The
profiles are not well fit by an r$^{1/4}$ or  linear
profile.  It is evident, however, that the slope of the monochromatic
profile (gaseous component) is steeper than the slope of the
continuum image (stellar component) outside a radius of 5 arcsec. This
is typical for gaseous disks in elliptical and lenticular galaxies.

\placefigure{fig03}

   A dust lane of $\sim$ 30-arcsec diameter crosses H16c from southeast
to northwest, at PA $\sim$ 120$^o$ (all PA measurements are made from
north to east). The galaxy contains patchy dust throughout its
center (within a radius of 10 -- 15 arcsec). Outside this radius the
subtraction of a concentric-elliptical isophote-model shows clean
residuals.  At large radii H16c has
non-concentric isophotes, whose centers are displaced in the direction
of H16d (Fig. 1).

\subsubsection{H16d}
   Dust is also detected in H16d, inside the central 15-arcsec radius.
In addition, H16d has a high surface brightness nucleus in the inner
2-arcsec radius (Fig. 2b).  A ``bar-like feature'' is visible out to
5$\arcsec$ and is embedded in dust.  These features are also present
but seem to be less prominent in the I image.  The effect of these
features can be noted in the ellipticity, PA and cos (4$\theta$) R and
I profiles shown in Fig. 4, for the inner region of the galaxy.  The
surface-brightness profiles in the R and I filters agree well except
for the cos (4$\theta$)-diagram.  Outside a radius of 8 arcsec the
galaxy is very flat (ellipticity=0.55).  The effect of the bright
nucleus and of the central bar-like structure in the inner 2 and 5
arcsec respectively is to cause an increase in the galaxy ellipticity,
a variation in the PA (with two discontinuities) and peaks in the cos
(4$\theta$) profile.  These peaks are much more prominent in the R than
in the I-band images (seen also in the cos (4$\theta$)-diagrams), due
to the effect of dust.  The very central region, shown in Fig. 4 
is also where the H$\alpha$ gas in emission is detected
(Section 3.2).  H16d has a second nucleus $\sim$ 7 $\arcsec$ east of
the main nucleus. This double nucleus is uncovered in a K
image of the galaxy (Mendes de Oliveira, unpublished).  
It is not visible in either the R or I bands, due
to obscuration by the dust.

  The R and I surface brightness profiles can be well described by a
r$^{1/4}$--profile  for radii greater than 8$\arcsec$, out to
61$\arcsec$ (at the lowest detectable isophote level of $\mu_R$ = 25
mag/arcsec$^2$). A fit to an r$^{1/4}$--profile gives an effective
radius for this galaxy of 10.5 arcsec (2 h$^{-1}$ kpc).  The ellipticity and PA
profiles are approximately constant for radii between 8$\arcsec$ and
61$\arcsec$ (not plotted) at values of 0.55 and 84$\degr$
respectively.

   H16d shows pointed isophotes, which, at small radii, could be due to
the central bar-like structure. However, this pointed morphology
dominates the whole structure of the galaxy, and not only its central
part, where the bar-like feature is detected.

\placefigure{fig04}

\subsection{H$\alpha$ morphology, kinematics of the gaseous disks and 
rotation curves}

\bf {The Data:}
\rm  Figs. 5 to 12 show the monochromatic images, the velocity fields
(VFs) of $H\alpha$ and the rotation curves (RCs) for galaxies H16a,b,c
and d.  The rotational velocities plotted have not been adjusted by the
cosmological correction  ({\it~1~+~z~}).

 Detailed discussion on the method used to obtain the systemic
velocity, center, PA of the major axis and inclination of the galaxy
are given in \cite{amr96}.  Advantages of Fabry-Perot
observations in the study of galaxy kinematics are also discussed in
that paper and in \cite{sch93}.

\subsubsection{H16a}

The pattern of the VF of H16a plotted in Fig. 5 is regular. The grand
design of the isovelocity lines describe a normal rotating disk with
differential rotation.  Morphologically, H16a is a barred galaxy
and the signature of the
bar is visible in the central parts of the VF.  The classical
distortions of the isovelocity lines when crossing a spiral arm are
clearly seen all along the arms.

 The RC of H16a shown in Fig. 6 was plotted taking into account all
velocity points except those within $20\degr$ of the galaxy minor
axis.  The cloud of small points represents the portion of measured
velocities within $\pm30\degr$ of the major axis in the plane of the
sky.  This gives a fair idea of the quality of our data. 
The mean RC of H16a (obtained by averaging both sides of
the galaxy) is flat from 2$\arcsec$ to the limit of our measurements
at $\sim$ 30$\arcsec$. In the outer 5$\arcsec$ 
small discrepancies between the two sides occur. These are
mainly due to the effects of the spiral arms.  The plateau is reached
within the inner 0.2 $h^{-1}$ kpc-radius.  The steep slope 
(fast rise) of the central RC is a typical signature of a bulgy galaxy.

  RHF91 found a peculiar gas velocity pattern for H16a. Outside
a radius of 16$\arcsec$, the rotation 
curve rises on one side of the galaxy and falls on the other.
These points are also plotted on Fig. 6. We do
not confirm this result. 

  We conclude that the RC of H16a is flat out to a 6.5 $h^{-1}$
kpc-radius (0.9 R$_{25}$) and that the mass distribution does not
seem perturbed by any interaction or merging in the group, inside this
radius.

\placefigure{fig05}

\placefigure{fig06}

\subsubsection{H16b}

  The $H\alpha$ emission in H16b is very weak and clumpy (see Fig. 7).  The
monochromatic map and the VF extend to a radius of $\sim 30 \arcsec$
(5.8 h$^{-1}$ kpc).  The total integrated flux is F(H$_{\alpha}$) =0.8
$\times$ 10$^{-14}$ erg s $^{-1}$ cm$^{-2}$.  The extended
emission-line gas lies in a disk of inclination 72$\degr$ $\pm$
3$\degr$ and position angle 70$\degr$ $\pm$ 3$\degr$.  
These parameters are used to
derive the RC plotted in Fig. 8.  This plot shows that the rotation
curve for H16b is non-axisymmetric. The northeastern side reaches a
minimum velocity of --73 km s$^{-1}$ while the southwestern side reaches
a maximum velocity of 173 km s$^{-1}$. In the central parts of the
galaxy the rotation curve rises very slowly 
compared to those of other spiral galaxies. 

RHF91 obtained a rotation curve for this galaxy, from measurements of
the H$\alpha$ and [NII] gas along the photometric major axis.  Their
results, completely different from ours, are overplotted on Fig. 8.
They found that the rotation is in the opposite sense of what we
measure. We have re-checked the raw data and have confirmed the sense
of rotation.  We cannot explain the discrepancy between the
results. The RCs  derived from the Fabry-Perot maps are based on the
average of many more measurements ($\sim$ 500 in this case) than for
the long-slit determinations. 

\placefigure{fig07}

\placefigure{fig08}

\subsubsection{H16c}

  The VF and the monochromatic image obtained for H16c were derived
from good signal-to-noise data (S/N $\sim$ 30--200), inside a
13$\arcsec$-radius.  As can be seen from Fig. 9a, the H$\alpha$
emission for H16c is strongly peaked in the central region, although it
can be detected out to a 30$\arcsec$-radius (6 h$^{-1}$ kpc).  The
H$\alpha$ flux goes from 32 $\times$ $10^{-14} erg s^{-1} cm^{-2}$
to 36 $\times$ $ 10^{-14} erg s^{-1} cm^{-2}$ if integrated within the
13$\arcsec$ or the 30$\arcsec$-radius.  Close to 90 $\%$ of the 
H$\alpha$-emitting 
gas mass of the galaxy is within the inner 13$\arcsec$ radius.

  The VF of H16c is reasonably regular, except for the presence of a
possible second component to the southwest of the galaxy (Section
3.4).  The analysis of the VF of H16c indicates that the extended
emission-line gas lies in a disk of inclination 60$\degr$ $\pm$
5$\degr$ and position angle 120$\degr$ $\pm$ 8$\degr$.  The morphology
of the line-emitting region is decoupled from that of the continuum-light
distribution (Fig. 9a).  Kinematic decoupling between gas and stars
is also observed (Fig. 9b and Section 3.3).

  The RC for H16c is plotted in Fig. 10a for a major-axis PA of
120$^o$, an inclination of 60$^o$ and the center measured from the VF.
The rotation curve resembles that of a gas disk in differential
rotation.

   The amplitude of the RC of the gas is similar for the receding and
approaching sides of the galaxy. The presence of a second gas
component does not significantly disturb the shape of the rotation
curve as described by the main body of the galaxy (when the center and PA
of the major axis are determined from the VF).

   In order to compare our results with the result of RHF91, we plotted
in Fig. 10b a RC with the center and PA derived from the R image. In
that study the photometric major axis of the galaxy was used to
position the long-slit.  It is clear that along the photometric major
axis of the galaxy the shape of the RC is ``sinusoidal'', mainly
because the second gas component is crossed.  Fig.  10b resembles the
RC determined by RHF91, from long-slit spectroscopy of the gas 
in emission.  A detailed discussion of Fig. 10 is in Section 4.1.

\placefigure{fig09}

\placefigure{fig10}

\subsubsection{H16d}

   The emission in H16d is much less extended than for the other three
galaxies and also strongly peaked.  The monochromatic map and the VF
(plotted in Fig. 11) extend to radii of $\sim$ 10 $\arcsec$
and $\sim$ 6$\arcsec$ respectively.  Analysis of the VF of
H16d indicates that the gas lies on a disk of inclination 47$\degr$
$\pm$ 5$\degr$ and position angle 70$\degr$ $\pm$ 8$\degr$.
These parameters are used to build the RC shown in Fig. 12.
This RC shows an extension of only $\pm$ 5$\arcsec$ on each side of
the galaxy.  The maximum velocity is 110 $\pm$ 7 km s$^{-1}$,
reached after 5 $\arcsec$ (1 h$^{-1}$ kpc).  
In the northeast side, the curve reaches a maximum
and then falls to about 60 $\pm$ 5 km s$^{-1}$.  This side of the
curve may be affected by the presence of a second nucleus, 7$\arcsec$ 
east of the bright central nucleus. In Fig. 11 the position of the
second nucleus is indicated in both pannels by a cross.

\placefigure{fig11}

\placefigure{fig12}

\subsection{Misalignment between gas and stars and kinematic warps}
 
   H16c displays the largest misalignment between the kinematic axis
inferred from the gas motions and the photometric (major) axis measured
from the stellar light (40$^o$)
followed by H16b (16$^o$) and H16d (14$^o$). For H16a the kinematic and
stellar axes are aligned.

   The major axis of H16c was measured from the H$\alpha$ image, the
VF, the continuum and R images (Table 2 contains derived quantities
for all four of the H16 galaxies).  The value of 115$^o \pm$ 10 $^o$,
determined from the monochromatic image, is very similar to that
defined by  the overall kinematics of the main body of the galaxy
determined from the VF (120$^o \pm$ 8$^o$, see Fig. 9b).  These are,
however, very different from the major axis defined by the stars, as
derived from the continuum, R and I images (80$^o \pm 5 ^o$), implying
a misalignment of $\sim$ 40$^o$ between the stellar and gaseous major
axes.

   H16c also has an offset between its gaseous kinematic center
and the bright optical center.  The kinematic center (shown with
a ``+'' in Fig. 9b) lies about 6$\arcsec$ (1.2 h$^{-1}$ kpc) to the
north of the brightest nucleus (also shown with a ``+'').  The
kinematic major axis is placed at the axis of symmetry of the main
kinematic body  (see Fig. 9b).  The center of the VF is chosen to be a
point along the major axis which makes the rotation curve symmetric
(with similar amplitudes for the receding and approaching sides).

  Misalignment between the kinematic and photometric major axes is also
detected for H16d, but in this case of only $\sim$ 14$^o$ (see Fig.
11b).  For H16b the major axis determined from the velocity field is
70$^o \pm 10$, different from the stellar major axis by $\sim$ 16$^o$
(see Fig. 7b).

   H16 b and c display strong kinematic warping, i.e.  variation of the
position angle of the kinematic major axis with radii. H16b presents a
60$\degr$ change of the PA of the major axis between a radius of 9$\arcsec$ and
27$\arcsec$. The kinematic major axis of the main component of H16c
shows a PA variation of over 60$\degr$ along the galaxy. No variation
of the PA of the major axis with radius was detected for H16 a and d.

\subsection{Possible second component in H16c}

  The 3900 $km s^{-1}$ iso-velocity contour in the southwest region of
the VF of H16c is not consistent with the rest of the map (see Fig.
9). This component is derived from profiles with S/N $=$ 90
pixel$^{-1}$. It is therefore not an artifact of the data.  In order to
show that this component is caused by noncircular gas motions we plot
in Fig. 13 the observed radial velocity at some given projected
distance from the kinematic center of rotation of the galaxy as a
function of the PA of the major axis. Plots for three values of
projected distance from the center, 5, 10 and 15 arcsec are shown.  A
clear deviation from a smooth curve occurs between PAs 220$^o$ and
280$^o$, due to the presence of a second gas component.

\placefigure{fig13}

\subsection{Velocity-dispersion measurements}

  The width of the emission-line
profiles of the galaxies are used to measure the velocity dispersion of the
ionized gas. 
   In the determination of the full-width of half maximum (fwhm)
of the emission-line profiles, we
checked if velocity gradients in the galaxies, low-S/N ratio of the
data, instrumental profile and seeing effects were artificially biasing
the results.  First the non-smoothed data were analyzed and the
instrumental profile (fwhm of $\sim$ 36 $km s^{-1}$) was used to
deconvolve the data.  Where the S/N was too low, some smoothing had to
be done and the process reiterated.  
At the distance of H16 one
pixel is a square of 200 $h^{-1}$ pc on a side. 

The fwhm of the profiles in the center of H16a is 125 $\pm$ 10 $km
s^{-1}$ and 80 $\pm$ 10 $km s^{-1}$ in the region of the spiral arms.
For H16c, the fwhm of the profiles in the central region is $\sim$ 130
$\pm$ 10 $km s^{-1}$ dropping quickly to $\sim$ 90 $\pm$ 10 $km s^{-1}$
to the east and south-east of the galaxy. For H16d, the fwhm is
$\sim$ 145 $\pm$ 15 $km s^{-1}$ in the central 5 arcsec and decreasing
quite rapidly but regularly to $\sim$ 30 $\pm$ 5 $km s^{-1}$ at 10" from
the center. These results are summarized in Table 2.
The low S/N for the H16b data cube did not allow the
determination of meaningful values of the fwhm for this galaxy.

\subsection{Mass determinations}

  The mass determinations are summarized in Table 2.  They were derived
for an adopted H$_o$ of 75 km s$^{-1} Mpc^{-1}$ and a distance to the
group of 51.35 Mpc.  We describe below how the values in Table 2 were
obtained.

  Treating non-circular motions as negligible perturbations, the mass
of a spiral galaxy out to a certain radius can be estimated from the
rotational velocity of the disk using the circular approximation.  We
estimated the total masses of the galaxies within a given radius
following \cite{leq83}.

  The ``warm'' (H$\alpha$-emitting) gas mass was obtained from the
formula given by Osterbrock (1974, case B recombination):  M$_{HII}$ =
2.8 $\times$ 10$^{-3}$ D$^2$ F(H$_{\alpha}$) n$_e$, where $D$ is the
distance to the group in Mpc, F(H$_{\alpha}$) is the total flux within
a given radius and n$_e$ is assumed to be 1000 cm$^{-3}$.  This may 
overestimate the gas mass, given that the value assumed for n$_e$
is an upper limit.  In the presence of dust, however, this  may
underestimate the warm-gas mass.

  The dust mass was obtained using the values from \cite{all96}
for the IRAS fluxes in 60 $\mu$ and 100 $\mu$.  We used the formula
from \cite{you89} to calculate the dust mass, with the
assumptions and approximations made by Roberts et al.  (1991, their
formula 7).

   The values for the X-ray emission in H16c and d are taken from
\cite{sar95}.  The formula (4) given by \cite{rob91}
was used to derive an X-ray gas mass for these galaxies.

  Based on independent measurements at several different
wavelengths, H16c and H16d contain significant amounts of 
interstellar matter.  The gas mass for H16c is about 6\% of the total
mass of the galaxy inside a radius of $\sim$ 0.9 R$_{25}$.

\section{General discussion of the results}

Each of the four HGC 16 galaxies shows various evidences for
interactions. This is clearly a very high density group. This gives an
excellent opportunity to correlate the different galaxy properties
suggested to be induced by galaxy-galaxy interactions. Via comparison
with the recent simulations of interacting galaxies, we can also
attempt to characterize the type of interaction each of the HCG 16
galaxies has undergone and to trace the interaction history of the
group. Before discussing these principal results of the study, we
address three other issues: the previous claims of unusual
rotation curves for these galaxies, how these galaxies fit
onto the Tully-Fisher and Fundamental Plane relations and the
presence of dwarf galaxies in the tidal tails of H16a.

\subsection  {Falling and/or sinusoidal rotation curves?}

   RHF91 presented evidence that a large fraction of the compact-group
spirals have falling rotation curves, unlike field spirals. If falling
rotation curves are common in compact groups, this would suggest that
the individual dark matter halos are absent most probably because they
were stripped by interactions.  In particular both H16c and H16d were
reported to have sinusoidal RCs, from measurements of the kinematics of
the gas obtained through long-slit spectra taken along the photometric
major axis of the galaxies.  Having the 2-D VFs of the galaxies we
can check on the reality of the sinusoidal shapes of the RCs.

   Fig. 10 shows two possible RCs for H16c (described in Section 3.2.3).
Fig.  10b was built with the same parameters used by RHF91, i.e.,
PA=78$\degr$ and the center coinciding with the optical/continuum
center.  This curve clearly has a sinusoidal shape, in agreement with
the curve presented in RHF91.

   The rotation curve of H16c plotted using the kinematical center,
inclination and position angle (Fig. 10a and see \S3.3 for
how these quantities are defined and derived) has a fairly normal shape
typical of a disk in differential rotation.  The sinusoidal rotation
curve derived through long-slit spectroscopy (Fig. 10b) along the
photometric major axis of H16c was caused by a misalignment between the
gas and stellar components and the presence of a second kinematic
component. 

There are a number of merger-candidates in the literature previously found to
have sinusoidal rotation curves. It would be
interesting to review these cases with a full 2-D velocity field 
analysis (e.g. \cite{mih98}, \cite{hib96}).

   In the case of H16b and H16d it is not possible to make their
rotation curves look normal by changing the center or the position
angle along which the curve is plotted. H16b has the peculiarity of
having one side (the northeast) with a much lower velocity amplitude
than the other side, while for H16d the velocity curve reaches a maximum and
drops on both sides.

\subsection{ The Tully-Fisher and the fundamental-plane relations}

   Dynamical processes may have worked very efficiently in HCG 16, as
exemplified by the high frequency of interaction indicators found for
all member galaxies.  It is of interest to investigate if the
``damage'' due to interactions and merging was large enough to affect
the position of the galaxies in the Tully-Fisher or fundamental-plane
relations.

   In order to check if H16a, H16c and H16d follow the TF relation we
have used the plots in Figs. 6, 10 and 12 to estimate the shape and
amplitude of the rotational velocity curves and we have then compared
these with ``template curves'' based on the TF relation for normal
galaxies (RHF91, Fig. 5).  No attempt is made to check if H16b follows
the TF relation since the two sides of its rotation curve do not
coincide.  H16a has a normal velocity amplitude for its luminosity,
following the TF relation.  For the main disk component of H16c the
measured V$_{max}$ is also within the expected value, for a galaxy of
this luminosity (see Fig. 5 of RHF91 and table 2).  A similar exercise,
for H16d shows that V$_{max}$ is $\leq 0.5$ of the value expected for a
normal galaxy of this luminosity, if the inclination given by the
velocity field (inclination = 47$\degr$) or R image
(inclination=57$\degr$) is used.  There is a chance, however that the
flattening of the velocity curve for H16d is due to dust. Alternatively
there could be problems in obtaining dynamical information about
gravity from the measured gas properties in the inner 5 $\arcsec$ of an
interacting galaxy. Dynamical information from the stars would be much
more reliable in this case.

   The surface-brightness of H16d follows an r$^{1/4}$-profile,
typical of early-type galaxies. It is, therefore, of interest to check if it
follows the r$_{eff}$ vs. $\mu_{eff}$ relation, a projection of the
fundamental-plane relation, observed for bulges and elliptical
galaxies.  With the values of r$_{reff}$ = 10.5 and $\mu_{eff}$ = 21.1
mag arcsec$^{-2}$, we find that H16d follows the fundamental plane
relations of normal field galaxies (\cite{kor89}).

\subsection{Dwarf galaxies in tidal debris ? }

  The analysis of R images of Hickson compact groups by \cite{hun96}
yielded a sample of 47 candidate dwarf
galaxies that may be associated with tidal arms and tails in
interacting and merging galaxies.  If the majority of these
dwarf-galaxy candidates are confirmed as being gravitationally bound
stellar systems, these authors estimated that a significant fraction
(perhaps as much as one-half) of the dwarf population in compact groups
is created in mergers occurred in the giant parent galaxies.

   Three dwarf candidates are suggested by \cite{hun96} to
sit on the eastern tail of H16a.  Two of the candidates were observed
in our fields.  We checked if emission lines are associated to the
candidates.  In both cases, the emission-line map has a S/N twice
higher on the corresponding pixels then in the direct neighbourhood of
these pixels. We therefore confirm these may be star-forming knots at
the redshift of the group. However, no conclusive information can be
obtained from the examination of the velocity field, the line-width and
the equivalent-line-width maps, due to insufficient S/N. We conclude
that the ``dwarf candidates'' are at the same redshift of the
group but we cannot exclude the possibility that they are material
connected to the main galaxies.  Longer exposure and higher spatial
resolution are necessary to confirm the nature of the emission, if
these are galaxies formed in interactions or just part of the main
galaxy which will fall back in a few crossing times. 

  It is interesting to note that these star-forming knots are found
exactly in the galaxy for which we show that no merging has
occurred (H16a).

\placetable{tbl-3}

\subsection{A census of the interaction-related properties for
the H16 galaxies}

There are a number of ``interaction indicators'' that have been suggested
over the years. These have been used to attempt to discern the different
types of interactions (primarily tidal interactions vs. merging)
that can occur between galaxies and are generally based on models
of colliding systems.

  Table 3 lists eight interaction/merging indicators and whether or not
each is seen in the four galaxies of H16.

  Close tidal encounters which do {\it not} lead to merging can
strongly affect the interstellar medium of the colliders, can trigger
strong bursts of star formation on a timescale of a few 10$^8$ years
but can only mildly affect the {\it kinematics} of the central velocity
fields of the colliding galaxies. In a merger, on the other hand, even
if the two nuclei have not yet mixed completely, the regular kinematic
structure of the disks disappear and the velocity fields may display strong
signatures of the collision (e.g.  the Antennae galaxies, \cite{amr92},
N7252, \cite{hib96}).  We, therefore, use the indicator
``highly disturbed velocity field'' as one of our definite evidences
for a merger.  Central double nuclei and/or double kinematic gas
components are other definite indicators of mergers. Double nuclei are
present, for example, in the well known mergers, the Antennae galaxy
(\cite{amr92}), N520 (\cite{hib96}) and IRAS 23128-5919
(\cite{mih98}) and double kinematic gas components are present
in IRAS 23128-5919, the Cartwheel galaxy (\cite{amr98}), HCG 31
and several other obvious merger candidates.

 We list as interaction indicators for collisions which do not
necessarily lead to merging the following properties:  kinematic
warping, gaseous and stellar major axes misalignments, tidal tails,
high infrared luminosity and central activity.  Although the presence
of tidal tails and plumes are common in mergers and they are often
taken as evidence for a major accretion event, we group it
with the indicators of mild interaction since they can also be formed
in interactions which do not lead to mergers.
Kinematic and photometric major axes misalignment and kinematic warping
are properties that are also usually associated with galaxy
collisions.  In a sample of 75 normal spiral galaxies studied by
\cite{sch93}, for which velocity fields were available, only
a very small fraction of the galaxies showed kinematic warping and/or
small gaseous/stellar major axes misalignments. Another indicator
listed in Table 3 is ``high infrared luminosities''. Certainly not all
strongly interacting galaxies have high IR luminosities but most IR
loud galaxies have disturbed morphologies.  Finally, ``central
activity'' is taken as an interaction indicator. Although there is not
a one-to-one correlation between interactions and central activity,
there are several lines of evidence that strongly suggest that
interactions may drive nuclear inflows and fuel central activity in
interacting systems.

\subsection {The interaction history of the H16 galaxies}

   The data presented in this paper support the existence of a regular
gaseous rotating disk in the center of galaxies H16a, b, c and d.  H16c
may contain two disks. Based on the interaction indicators for each
galaxy listed in Table 3 and in the context of the various models of
galaxy interactions (\cite{bar89}, \cite{bar92}, \cite{bar96})
we suggest the following histories for the H16
galaxies.

{\bf H16a:} This galaxy does not present any of the indicators common
to merging galaxies, listed in Table 3. 
It has a normal mass distribution inside a radius
of 30 arcsec and it does not contain any multiple nuclei or gas
components.  At large radii, however, the galaxy looks morphologically
disturbed and shows tidal tails to the east and south.  Although
numerical models of compact groups suggest that galaxies with extended
tidal tails are excellent candidate mergers (Fig. 1 of \cite{bar89}),
the grand-design isovelocity lines of H16a with no signs of
disturbances inside a radius of $\sim$ R$_{25}$ strongly suggest that
this galaxy has not yet suffered a major merger.

An interaction between galaxies H16a and H16b is very probable.  The
difference in systemic velocities between these two galaxies is less
than 200 km s$^{-1}$ and their isophotes overlap. 
The normality of the velocity field of this galaxy compared with
the peculiarities observed for the velocity fields of the other three
very kinematically disturbed companions, leads us to the conclusion
this may be a fairly new member to the heart of the group.

{\bf H16b:} Although H16b is classified as an Sb galaxy, it has a
velocity field which is very peculiar, unlike that observed for any
other normal spiral galaxy (e.g. \cite{sch93}).  Strong warp
of the kinematic major axis, misalignment between the optical and
kinematic axes and a very slow rise in the velocities in the inner
regions of the galaxies are obvious peculiarities related to past or
ongoing interactions.  In addition, the blue and red sides of its
velocity curve do not match in shape nor in amplitude and its visual
morphology is asymmetric, although smooth and ``well behaved'' (if
compared with the morphologies of other merger remnants).  The close
proximity of H16a and the overlapping isophotes with this galaxy
suggest a clear ongoing interaction. H16b has little ionized gas
compared to the other members of the group studied here (Table 2).
Also the mass of molecular gas is four times smaller than the average
value for the other members.  It is the one galaxy from the four
studied here that is not detected at radio or infrared wavelengths.
Although H16b does not show a double nucleus or a double gas system,
which would be a clear sign of an ongoing merger, the very disturbed
velocity field calls for a scenario where strong interactions or
accretion events have occurred in the past, which have, however, left
the disky morphology of the galaxy intact.  A gas-poor accretion event
is possible, given the lack of gas in H16b. The scarcity of gas and
consequent 
low S/N of the data did not allow detection of a double gas component,
if it was present.  In addition to having suffered accretion
events H16b may be involved in an on-going interaction with H16a,
which may result in a major merger.  Unlike the two ongoing mergers in
H16 (H16c and H16d, see below) the warm gas content of H16b is
not concentrated to the center of the galaxy.

{\bf H16c:} This galaxy is clearly an ongoing starburst merging
system.  It presents seven indicators (all but ``tidal tails'') for
strong interactions and mergers, according to Table 3.  The VF of H16c
shows the existence of a second velocity component to the southwest of
the galaxy, closely spaced in velocity and position to the main
component.  The observed kinematic warp  and misalignment between gas
and stars are strong evidences for external accretion of the central
ionized gas in H16c. Its infrared luminosity is comparable to that of
other mergers (log L$_{IR}$ = 11.6 L$\odot$). The morphology of the
galaxy as a whole is amorphous.  No disks or spiral arms or tails are
seen down to a surface brightness limit of R $=$ 25 mag arcsec$^{-2}$.
The shapes of the continuum and monochromatic images of H16c (Fig. 4)
indicate that the emission-line isophotes are significantly flatter
than the isophotes of the underlying stellar component, typical of
early-type galaxies. However, the light profile of the galaxy does not
follow an r$^{1/4}$ law as is common for relaxed mergers.

The kinematics of H16c resembles that of the merger remnant N6240
(\cite{bla91}). The overall optical morphology of H16c,
however, is much less peculiar than that of N6240 or of any other ongoing
merger remnant from the Toomre and Toomre's sequence.  It is possible
that pre-existing tidal arms or tails were quickly ripped apart by the
dynamical forces within the compact group, a situation which is 
unlikely to happen in less dense environments (\cite{bar92}).
Alternatively tidal tails may have never been formed.

   The HI map of the group (Williams, private communication) shows
a strong connection between galaxies H16c and H16d. It also shows that
the centers of the HI contours are displaced from the optical centers
of H16c and H16d by almost one optical radius, in the direction of
the line that joins the centers of these two galaxies. 

{\bf H16d:} This galaxy may also be a merger remnant.  It presents five
of the eight indicators of interactions and mergers, from Table 3.
Double kinematic gas component, kinematic warping and tidal tails are
not detected for this galaxy.  The first two indicators are not
observed probably due to the small range in radii where the gas
kinematics can be measured. The lack of tidal tails is common to all
mergers in group H16 and may be a common feature of mergers in compact
groups in general.

   H16d has a double nucleus and a peculiar velocity field.  It has a
high infrared luminosity (log L$_{IR}$ = 11.7 L$\odot$), similar to that
of other mergers.  Outside a radius of 8 arcsec the light profiles of
the galaxy (in the R and I bands) follow an r$^{1/4}$-profile with an
r$_{eff}$ of 10.5 arcsec.  In the inner parts the galaxy profiles
flatten, as is typical of the mass profiles of modeled merger remnants
(eg. \cite{brn92}).  H16d has a peculiar bar-like structure in the
central 10$\arcsec$ which coincides with the peaked ionized gas,
molecular gas and dust components.  The rotation curve of H16d peaks
and falls on both sides within the inner 2 h$^{-1}$ kpc region. We do
not have kinematical information on the second nucleus, which lies 7
arcsec to the east of the main nucleus. The second nucleus is only seen
in the NIR due to obscuration from dust in the optical.  The H$\alpha$
emission in H16d (and in H16c also) is strongly spatially concentrated,
in qualitative agreement with the results of simulations of gaseous
collisions, in which the gas is driven towards the nucleus of the
galaxies (e.g. \cite{bar96}).

   Our measurements support the models of dynamical evolution of
compact group galaxies (\cite{brn92}) and should impose strong
constraints to the detailed simulations of the gas kinematics during
group merging. \\

\acknowledgments

The authors thank Barbara Williams for providing an HI map of H16,
Chantal Balkowski, Roger Coziol and Jacqueline van Gorkom  for
insightful discussions and J.L. Gach for helping during the
observations.  CMdO acknowledges the financial support from the
Alexander von Humboldt foundation.  H. Plana acknowledges the financial
support of the Brazilian FAPESP, under contract 96/06722-0.

\clearpage
 
\begin{deluxetable}{lrr}
\tablecaption{Journal of Perot-Fabry observations
\label{tbl-1}
}
\tablewidth{0pc}
\tablehead{
\colhead{} & \colhead{Compact Group of Galaxies HCG 16} & \colhead{} 
}
\startdata
Observations        & Telescope  & ESO 3.6m \nl
                & Instrument & CIGALE @ Cassegrain focus \nl   
               & Date       & August 26$^{th}$ and 27$^{th}$ 1995 \nl   
                & Seeing     & $\sim$1" \nl  
Interference Filter & Central Wavelength & 6653 \AA \tablenotemark{1} \nl
                    & FWHM               & 24~\AA \tablenotemark{2}  \nl
                    & Transmission & 0.70/0.69/0.65/0.65\tablenotemark{1,3}  \nl
Calibration         & Neon Comparison light & $\lambda$ 6598.95~\AA \nl
Perot-Fabry         & Interference Order & 796 @ 6562.78 \nl
                    & Free spectral range at H$\alpha$ (km s$^{-1}$) & 387.7 \nl
                    & Finesse & 12 at H$\alpha$ \nl
                    & Spectral resolution at H$\alpha$ & 9600 \nl
Sampling            & Number of Scanning Steps & 24 \nl
                    & Sampling Step   & 0.36~\AA (16.15 km s$^{-1}$) \nl
                    & Total Field & 230" $\times$ 230" (256 $\times$ 256 px$^2$) \nl
                    & Pixel size & 0.91" \nl
Detector & Photon Counting Camera (IPCS) & Time sampling of 1/50 s \nl
Exposures times & Total exposure & 2 hours per field \nl
& Elementary scanning exposure time & 5 s per channel \nl
& Total exposure time per channel & 300 s \nl 
\enddata

\tablenotetext{1} { For a temperature of 5$\degr$}
\tablenotetext{2} { For a mean beam inclination of 2.7$\degr$ }
\tablenotetext{3} { Transmission for the mean redshift of H16a,b,c and d
respectively}

\end{deluxetable}
 
\clearpage

\begin{deluxetable}{lrrrr}
\scriptsize
\tablecaption{General and physical parameters for the H16 galaxies \label{tbl-2}}
\tablewidth{0pc}
\tablehead{
\colhead{Name} & \colhead{HCG 16a} & \colhead{HCG 16b}   & \colhead{HCG 16c}   &
\colhead{HCG 16d}
} 
\startdata
Other Names & N835/Mrk 1021 & N833 & N838/Mrk 1022 & N839 \nl

$\alpha$ (1950)\tablenotemark{1} &  02$^{h}$06$^{m}$57.4$^{s}$&  
02$^{h}$06$^{m}$53.3$^{s}$ 
 & 02$^{h}$07$^{m}$11.3$^{s}$ &  
02$^{h}$07$^{m}$15.6$^{s}$\nl

$\delta$ (1950)\tablenotemark{1} & --10\degr22'20.0" &  --10\degr22'08.9"& --10\degr22'56.0" & 
--10\degr25'11.1" \nl

Morphological type \tablenotemark{1} \tablenotemark{,} \tablenotemark{2}  
& SBab/SBab & Sab/Sa & Im/SO(pec) &Im/SO(pec)\nl

Systemic Velocity (km s$^{-1}$) \tablenotemark{1} \tablenotemark{,} \tablenotemark{3} & 4152/4052 & 3977/3837  & 3851/3858 & 3847/3864\nl

D(Mpc) H$_o$ $=$ 75 $km s^{-1} Mpc^{-1}$ 
& 51.35 & 51.35 & 51.35 & 51.35 \nl

B$_{T}$ (mag) \tablenotemark{1}  & 12.7 & 13.2  & 13.1 & 13.4 \nl

D$_{25}$/2 (") \tablenotemark{2} 
& 37$\pm$10 
& 45$\pm$7
& 34$\pm$10 
& 43$\pm$4\nl 

\tableline

\multicolumn{5}{l}{
Fluxes and masses of the different ISM components
}\nl

\tableline

F(H$\alpha$)10$^{-14}$ erg s$ ^{-1}$ cm$^{-2}$ & 12.8 & 0.8 & 36.0 & 20.0 \nl

Log L(H$\alpha$) erg s$ ^{-1}$ & 39.5 & 38.3  & 40.0 &  39.7 \nl

M$_{HII}$ (10$^{4}$ M$_{\odot}$) & 9.4 & 0.6  & 26.5 & 14.7 \nl

M$_{tot}$ (10$^{10}$ M$_{\odot}$)/R$_{max}$ (h$^{-1}$kpc) & 8.3/6.3 & 4.3/6.7 & 7.2/7.5 & 0.23/1.1 \nl

M$_{X}$ (10$^{8}$  M$_{\odot}$)                & 28.8\tablenotemark{4} &    & 9.1 & 6.6\nl

M$_{H_{2}}$ (10$^{9}$  M$_{\odot}$)\tablenotemark{5}  & 2.4 & 0.6 & 2.6 & 1.9\nl

M$_{Dust}$ (10$^{7}$  M$_{\odot}$) & 1.07 & $<$ 0.04  & 1.5 & 1.7\nl

\tableline

\multicolumn{5}{l}{
Major-axis position angles (this study)
}\nl

\tableline

Velocity Field  & 3\degr $\pm$ 10\degr & 70\degr $\pm$ 3\degr & 120\degr $\pm$ 
8\degr & 70\degr $\pm$ 8\degr \nl

Monochromatic Map & 0\degr $\pm$ 10\degr & 86\degr $\pm$ 10\degr  & 115\degr 
$\pm$ 10\degr &  80\degr $\pm$ 7\degr\nl

Optical Image & 3\degr $\pm$ 5\degr & 86\degr $\pm$ 10\degr & 80\degr $\pm$ 5\degr  & 84\degr $\pm$ 5\degr \nl

\tableline
\multicolumn{5}{l}{
Inclination of the disk (this study)
}\nl
\tableline

Velocity Field  
&  43\degr $\pm$  5\degr & 72\degr $\pm$ 3\degr  & 60\degr $\pm$ 
5\degr & 47\degr $\pm$ 5\degr 
\nl
Continuum image  & 57\degr $\pm$ 5\degr & 58\degr $\pm$ 5\degr  &
48\degr $\pm$ 8\degr & 27\degr $\pm$ 7\degr \nl
R image (at R$_{25}$)  & 38\degr $\pm$ 8\degr & 66\degr $\pm$ 8\degr & 44\degr $\pm$ 8\degr & 57\degr $\pm$ 5\degr \nl

\tableline
\multicolumn{5}{l}{
Parameters from the gas VF and Rotation Curves (this study)
}\nl
\tableline

Systemic Velocity (km s$^{-1}$) & 4040  & 3905 & 3850 &  3845 \nl
Min. velocity (projected, km s$^{-1}$) & --163 & --77 & --220 & --78 \nl
Max. velocity (projected, km s$^{-1}$) & 174 & 183 & 225 & 76 \nl 
FWHM of central profiles (km s$^{-1}$) & 125 &  & 130 & 145 \nl

\enddata

\tablenotetext{1}{ from \cite{hic93}}
\tablenotetext{2}{ from \cite{dev91}}
\tablenotetext{3}{ from RHF91}
\tablenotetext{4}{ mass is given for H16A+B}
\tablenotetext{5}{ from \cite{bos96}}

\end{deluxetable}

\begin{deluxetable}{lrrrr}
\tablecaption{Interaction indicators
\label{tbl-3}
}
\tablewidth{0pc}
\tablehead{
\colhead{} & \colhead{HCG 16a} & \colhead{HCG 16b} 
& \colhead{HCG 16c} & \colhead{HCG 16d}
}
\startdata
Highly disturbed velocity field\tablenotemark{1} & -- & $+$  & $+$ & $+$ \\
Central double nuclei\tablenotemark{1}   & -- & --  & $+$ & $+$ \\
Double kinematic gas component\tablenotemark{1} & -- & --  & $+$ & -- \\
Kinematic warping\tablenotemark{2} & -- & $+$  & $+$ & -- \\
Gaseous {\it vs.} stellar major axis misalignment\tablenotemark{2} & -- & $+$  &
$+$ &
$+$ \\
Tidal tails\tablenotemark{2} & $+$ & --  & -- & -- \\
High IR luminosity\tablenotemark{2} & $+$ & -- & $+$ & $+$ \\
Central activity\tablenotemark{2} & $+$ & $+$ & $+$ & $+$ \\
\enddata

\tablenotetext{1} { Indicator definitely associated with mergers}
\tablenotetext{2} { Indicator does not require that a merger has occurred}

\end{deluxetable}

%
%


\clearpage

%
%

\figcaption[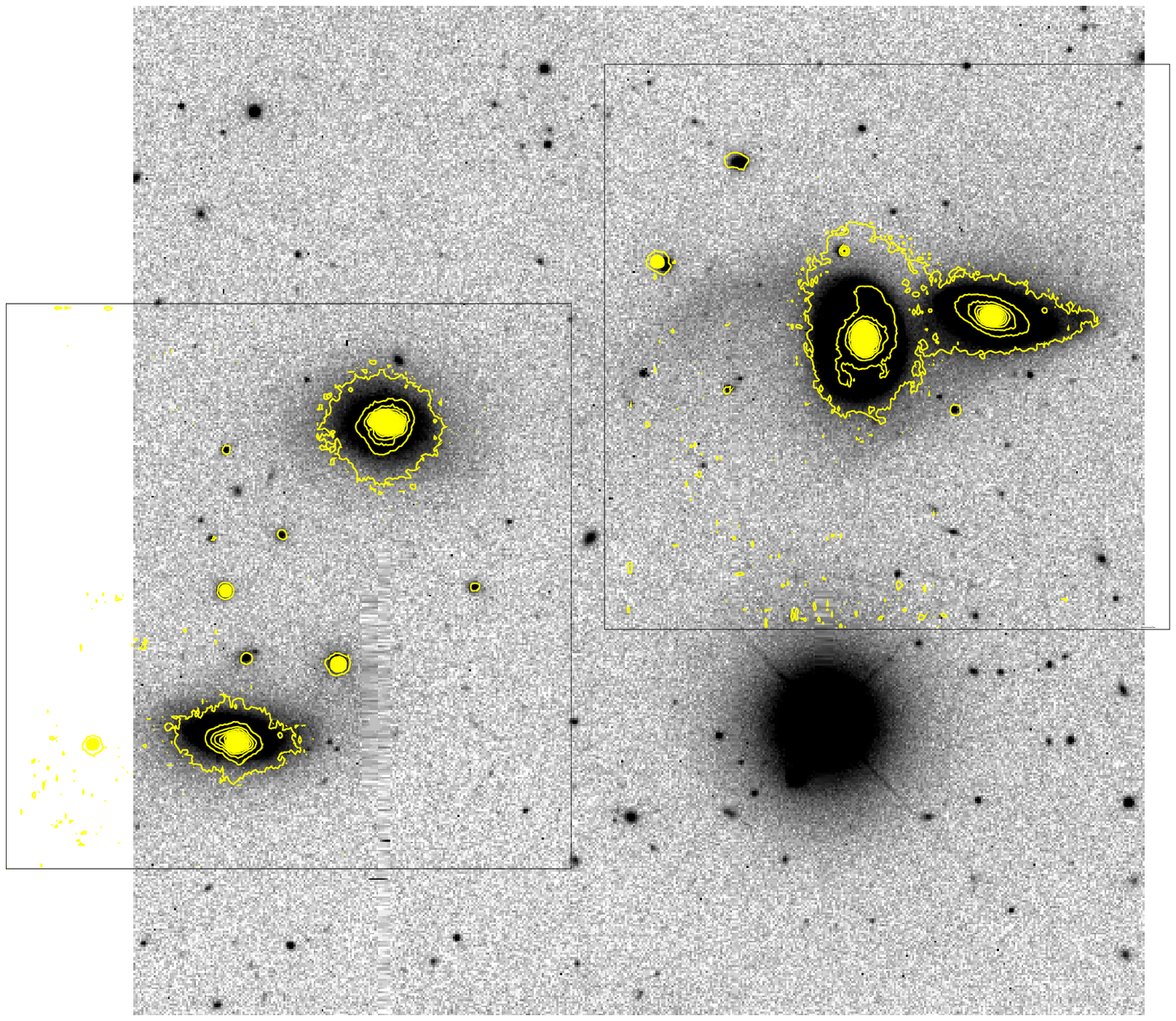]{
Continuum contour plots superimposed on the I image of H16.
North is up; east is to the left.  The two
squares (230" on a side) indicate the field of view of the
Fabry-Perot instrument for the two observed frames.  The I-image is 7.7'
on a side. The continuum contours are plotted in a linear arbitrary scale.
\label{fig01}}

%
%

\figcaption[fig2.ps]{
R image of the inner regions of H16 c and d.  The box around each image
is 12-arcsec on a side.
North is up and east is to the left. 
The center of the
concentric-elliptical isophotes outside a radius of 10 arcsec is
placed in the middle of the box. Elliptical isophotes with
a major-axis radius of 5 arcsec ($\sim$ 1 h$^{-1}$ kpc) 
are draw on the figures for
guidance on the scale only.  Left image, H16c: 
the double nucleus is not
coincident with the center of the main body of the galaxy, as described
by its outer isophotes and kinematics.  Right image, H16d:  the
galaxy has an inner high surface brightness nucleus with
a bar-like feature within its inner 5" radius.
\label{fig02}
}

%
%
\figcaption[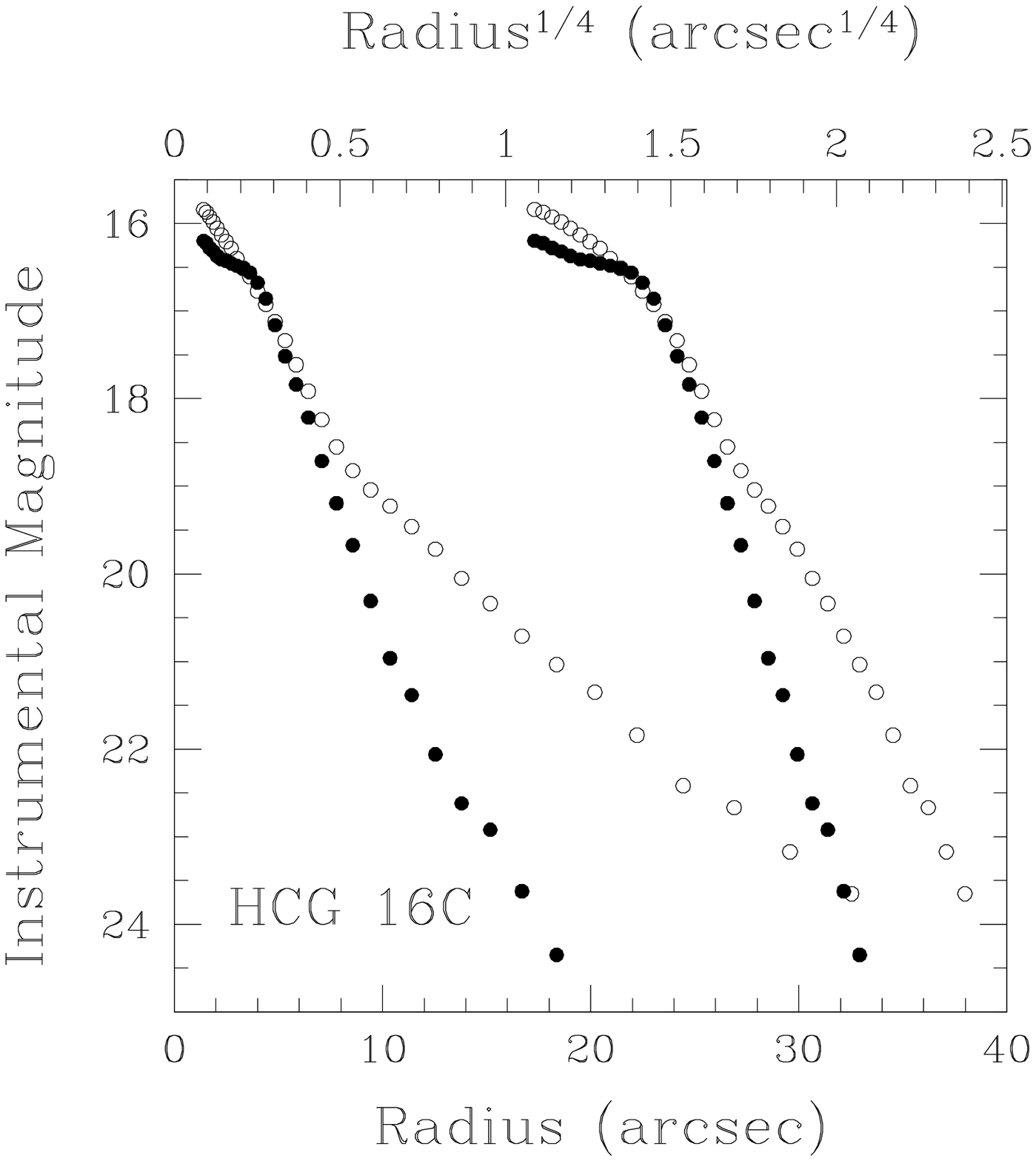]{
Surface brightness profiles of H16c obtained for the gas component
(closed circles), and stars (open circles).  The top label and axis
correspond to the two curves on the right of the plot (fits to an
r$^{1/4}$--profile) while the bottom label and axis relates to the two
curves on the left of the diagram (linear fits). The vertical
normalization is done arbitrarily.  The profiles are not well fit by an
r$^{1/4}$ or  linear profile.  It is evident, however, that the slope
of the monochromatic profile (gaseous component) is much steeper than
the slope of the continuum image (stellar component) outside a radius
of 5 arcsec. This is typical of gaseous disks in elliptical and
lenticular galaxies. \label{fig03}} 

%
%
\figcaption[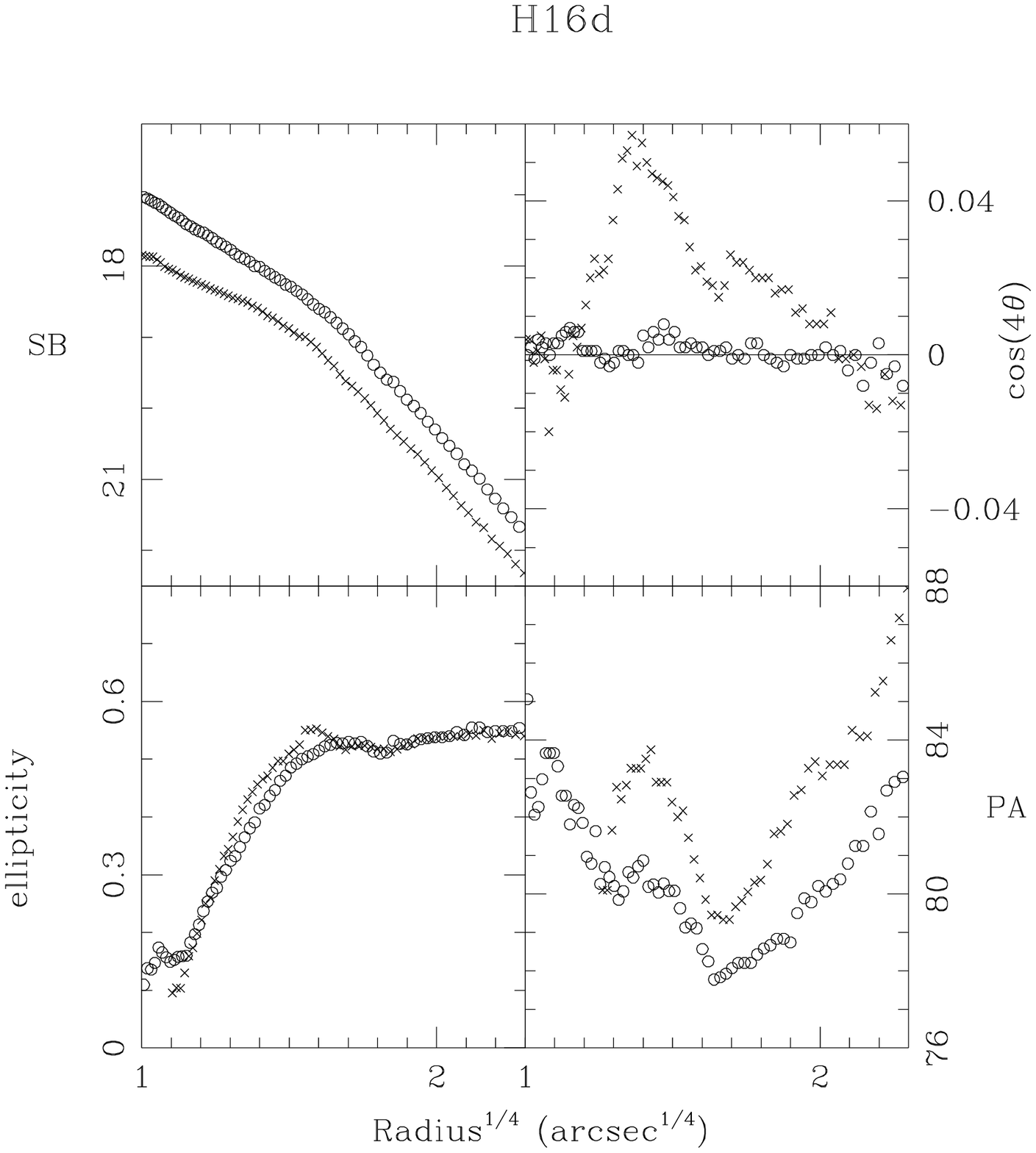]
{Surface brightness, ellipticity, position angle and
cos (4$\theta$) profiles for H16d in R (X) and I (open circles)
See text for details.
\label{fig04}}

%
%
\figcaption[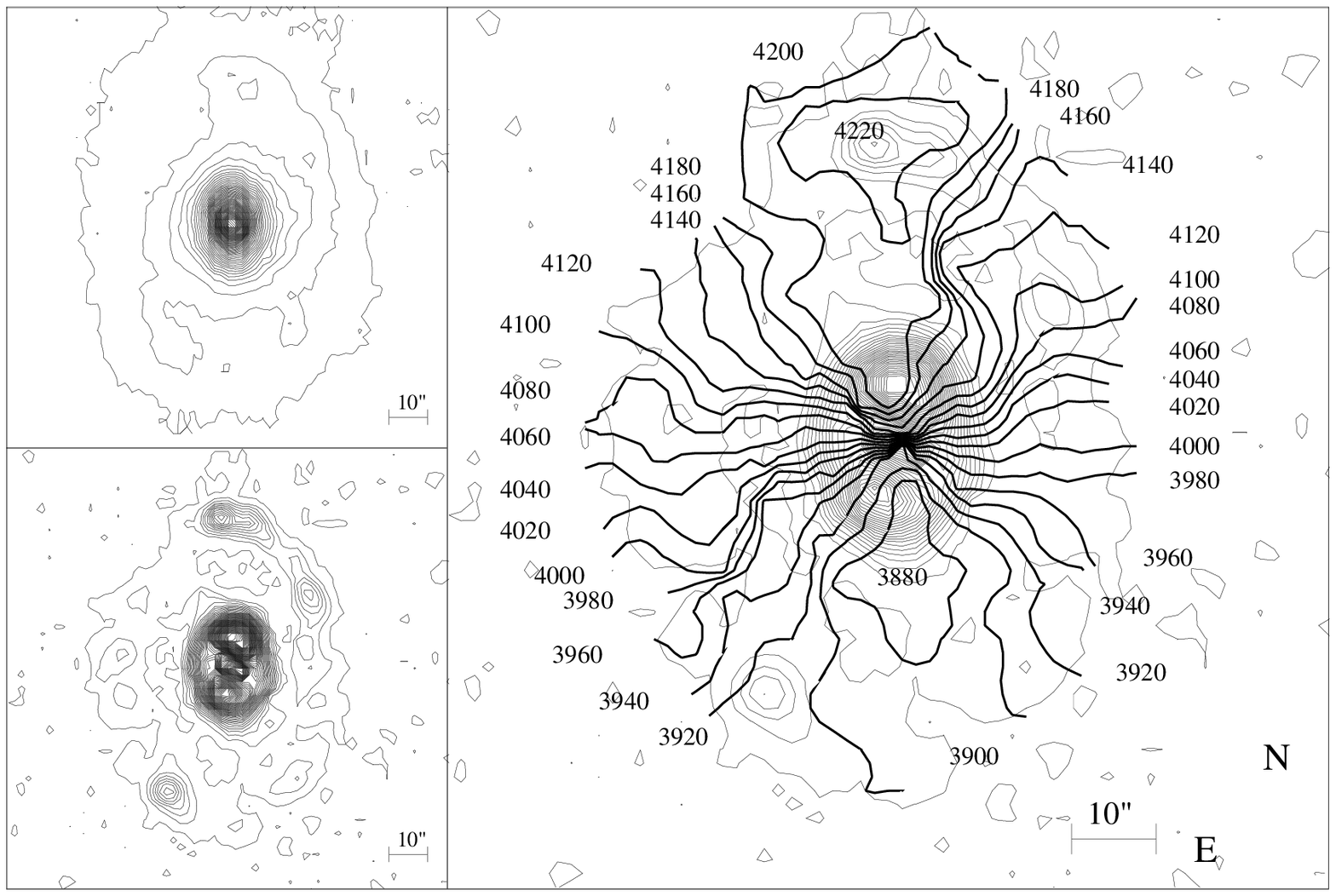]{
H16a -- 
Upper-left: continuum iso-intensity contours in a linear arbitrary scale.
Bottom-left: monochromatic iso-intensity map, calibrated
in units of 10$^{-17} erg s^{-1} cm^{-2} arsec^{-2}$. 
The lowest level is 0.65 and the step is 2.6.
Right pannel: 
Isovelocity contours of the ionized gas (thick solid lines) are
superimposed on the monochromatic image (thin solid lines).
The heliocentric radial isovelocity lines have been drawn after
smoothing of the original profiles
with a gaussian function (of fwhm 1.5''~$\times$ 1.5'').
\label{fig05}}

%
%

\figcaption[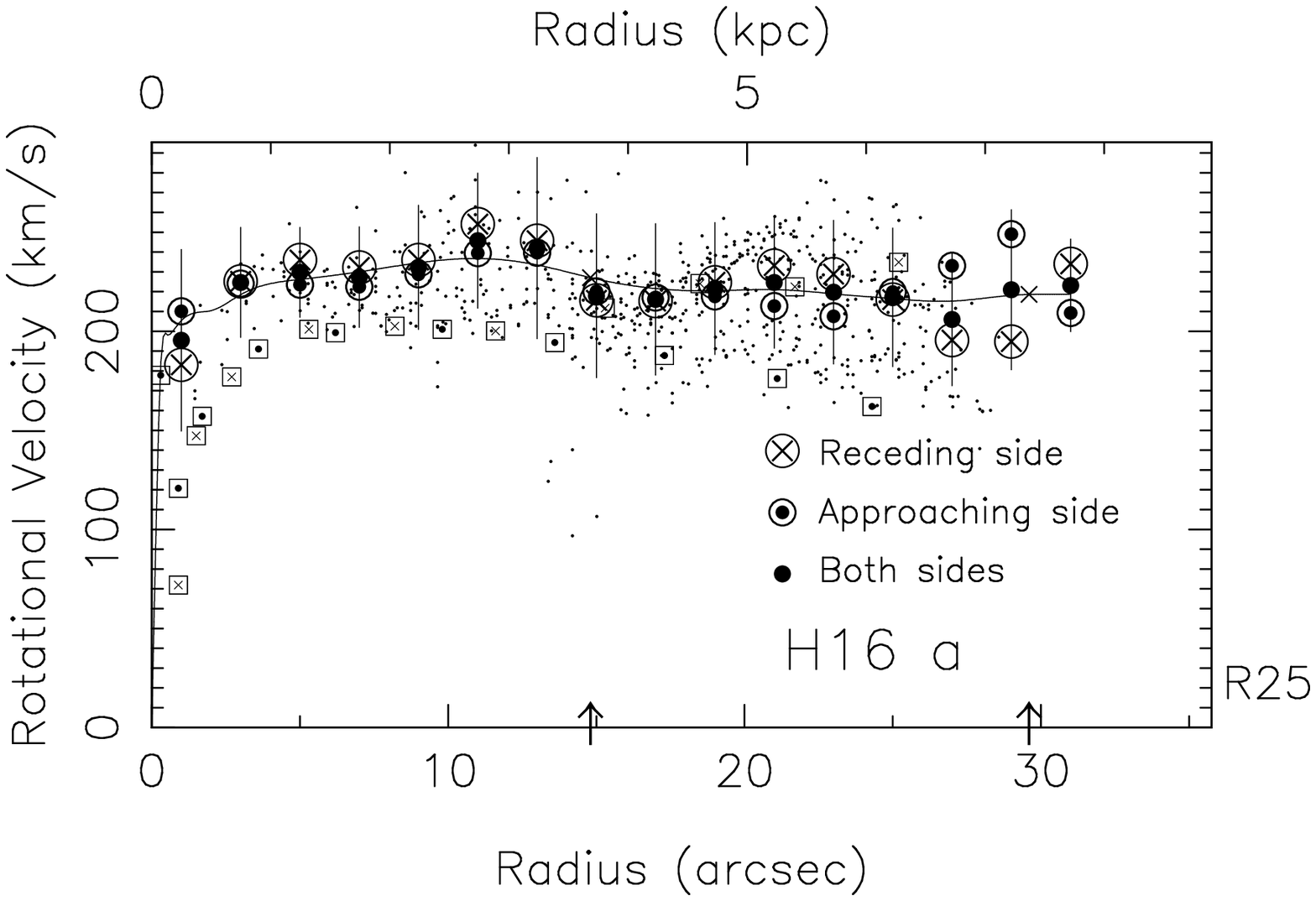]{
Rotation curve obtained from both sides of H16a.  The cloud of points
represents the portion of measured velocities within $\pm30\degr$ of
the major axis in the plane of the sky.  The large velocity points are
weighted averages for the entire data set (within $\pm70\degr$ of
the major axis PA) over successive concentric annuli in
the plane of the galaxy.  The weights are proportional to cos $\theta$,
where $\theta$ is the angular separation of a pixel from the major axis
as measured in the plane of the galaxy. For each annulus we give the
average velocity for the receding side ($\otimes$), the approaching
side ($\odot$), and the average of both sides (small filled circle),
with a $\pm1$ rms error bar. Points enclosed within parentheses () are
derived from only one or two measured velocities. The final rotation
curve (thick solid line) is a cubic spline function least-square fitted
to the average points, weighted by their errors.  The large squares
($\Box$) plot the rotation points of RWF adopting our inclination (The
size of the square is roughly the mean error adopted by RHF91).
The receding side is showed by an ``X'' within the square and the
approaching side by a point within the square.  A value of
$H_{0}~=~75~km~s^{-1}~Mpc^{-1}$ has been adopted for the distance
scale.  The arrows along the abscissa indicate the $0.4R_{25}$ and
$0.8R_{25}$.
Small crosses are plotted on
the fitted RC at these abscissae.  $R_{25}$ is the corrected radius
at the 25th~magnitude~arcsec$^{-2}$ isophote, as given in Table 2
(D$_{25}$/2).
\label{fig06}}

%
%

\figcaption[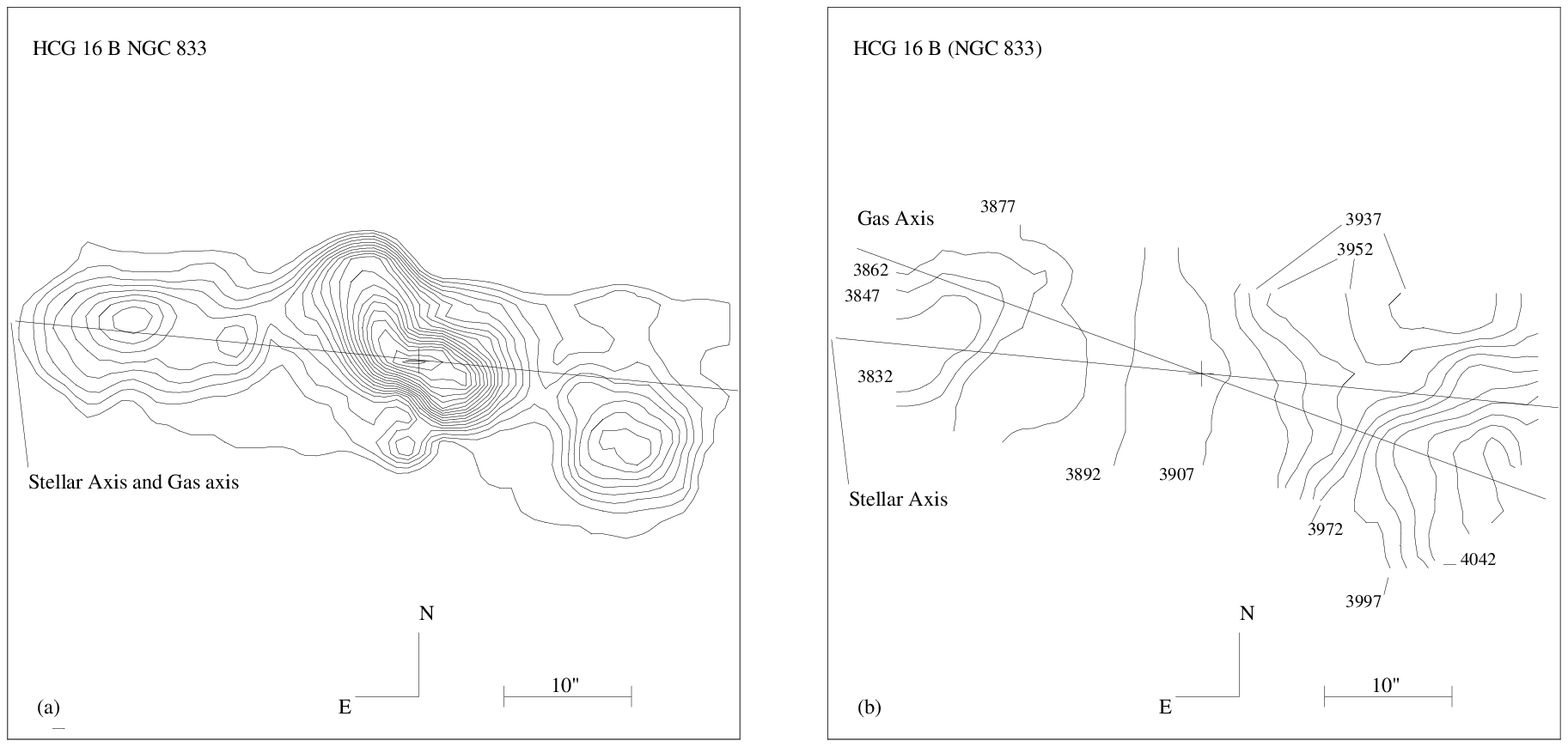]{
(a) Monochromatic image for H16b, calibrated in units of 10$^{-18}$ erg
s$^{-1}$ cm$^{-2}$ arcsec$^{-2}$.  The lowest level is 3 and the step
is 1.5. The continuous line represents the stellar major axis which is
coincident with the major axis defined by the morphology of the gas.
(b) Velocity field built from the H$\alpha$ line emission.  The two
continous lines represent the stellar major axis (same as in Fig. 7a)
and the major axis defined by the kinematics of the gas.
\label{fig07}}

%
%

\figcaption[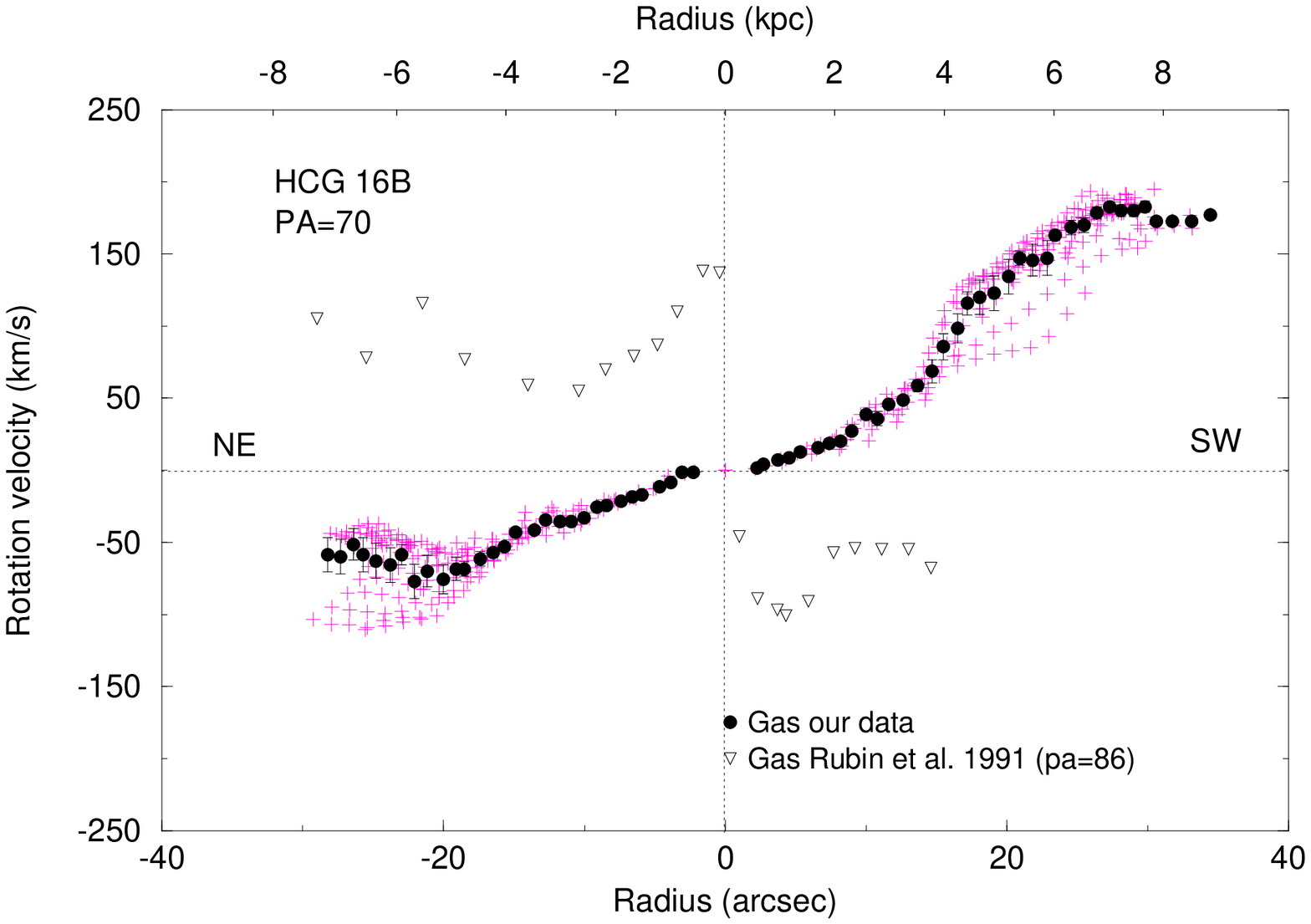]{
Rotation curve for H16b (PA=70$\degr$).  The crosses are weighted
averages within 30$\degr$ of the kinematic major axis PA.
Filled circles represent the mean radial velocities. RHF91's data are
overplotted as open triangles.\label{fig08}}

%
%

\figcaption[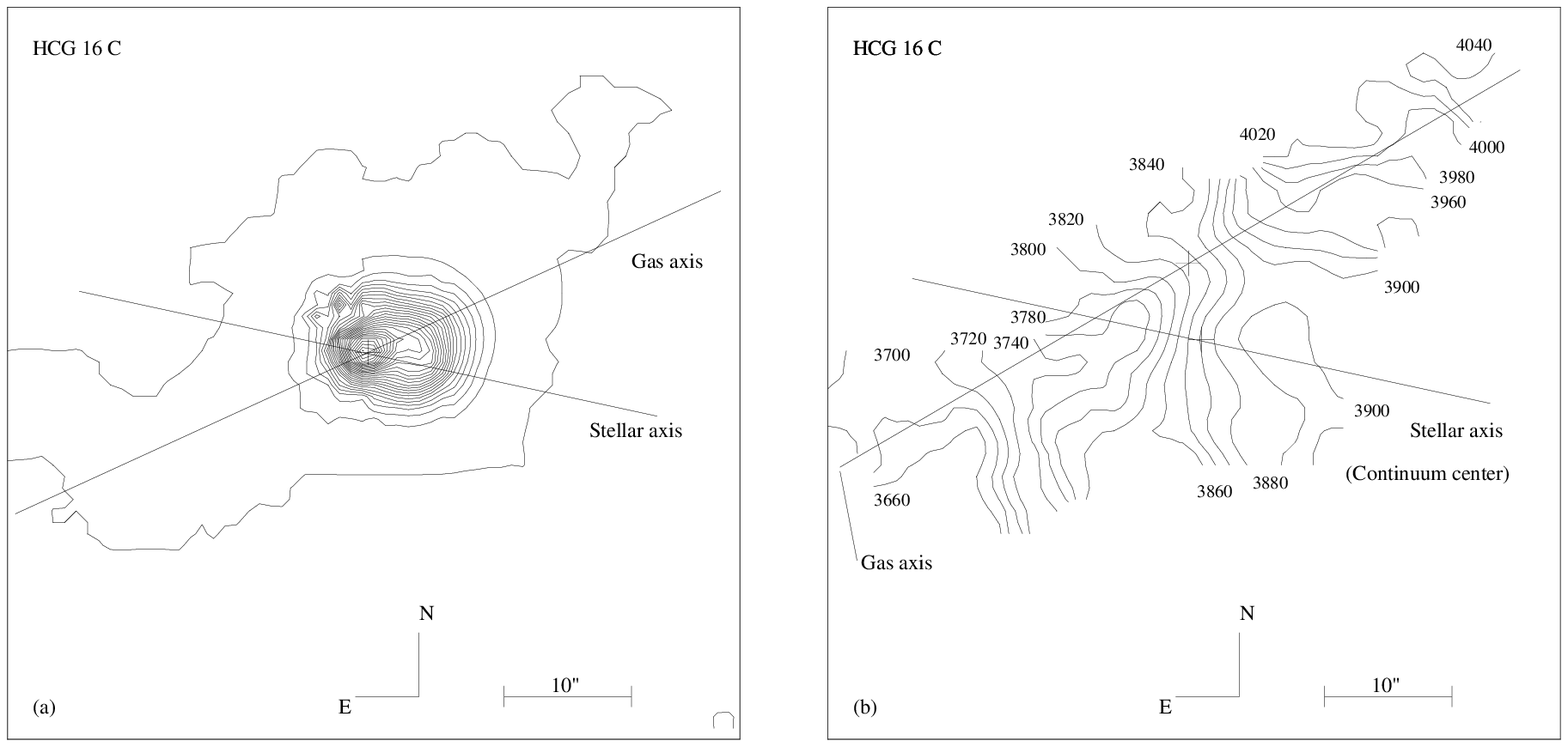]
{(a) The monochromatic iso-intensity map for H16c, calibrated
in units of 10$^{-17} $erg/s/cm$^2$/arsec$^2$. The lowest level is 0.7
and the step is 20. The two continuous lines, represent the stellar (SA)
and gas (GA) major axes. The intersection of these two lines mark the
monocromatic center.  (b) The velocity map built from the
H$_\alpha$-line emission.  The upper and lower crosses represent the
kinematical and continuum centers respectively.\label{fig09}}

%
%

\figcaption[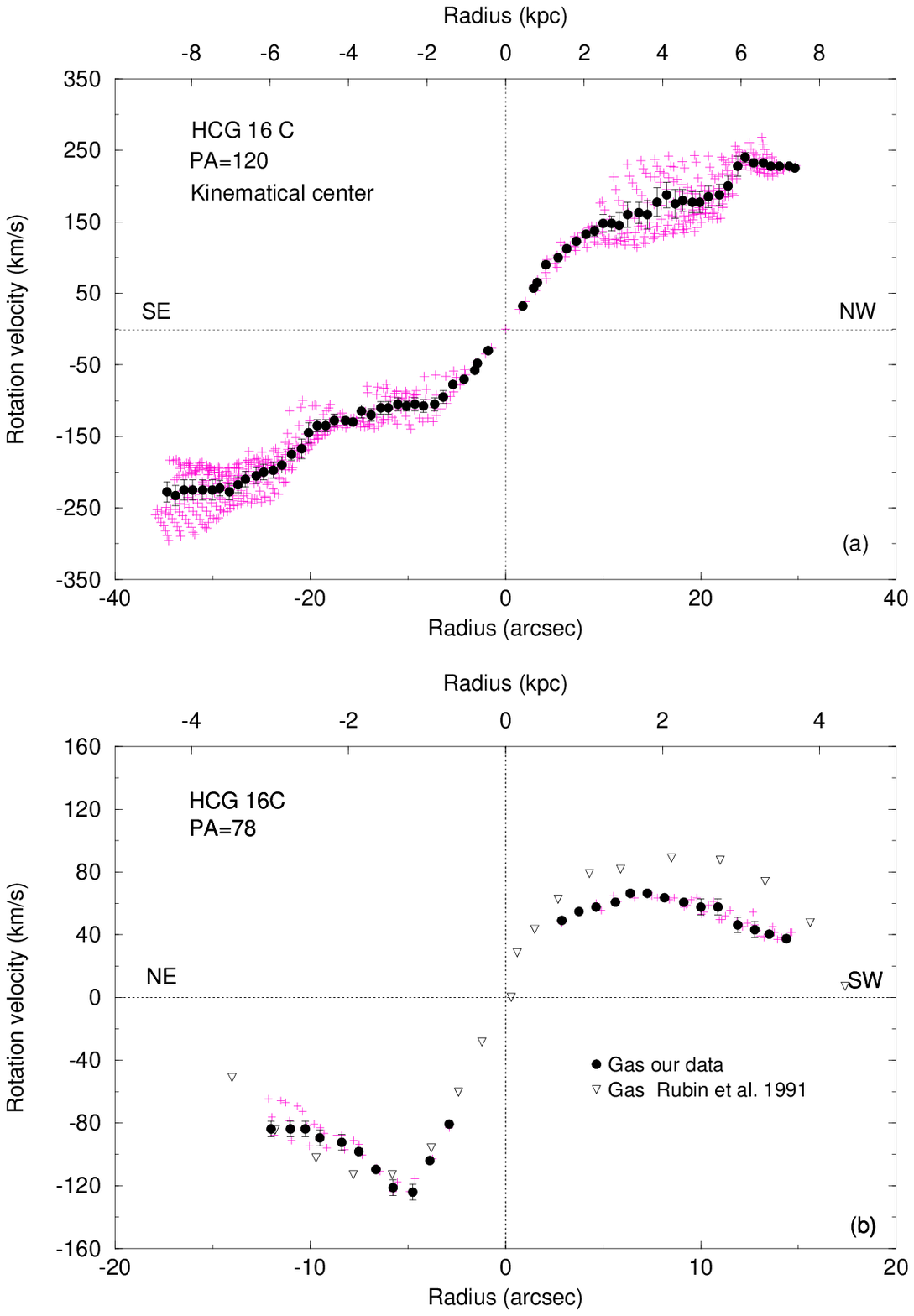]{Rotation curve for H16c:
(a) Crosses represent all measured values within 40\degr of the
kinematic  major
axis (PA=120$\degr$). The center is taken
to be the kinematic center. The solid dots represent the average
velocities. (b) Crosses and solid dots as in Fig. 10a, but for
a major axis with PA=78$\degr$ and with
the continuum
center. Overplotted are the velocities measured by RHF91 as open 
triangles. See text for details. 
\label{fig10}}

%
%

\figcaption[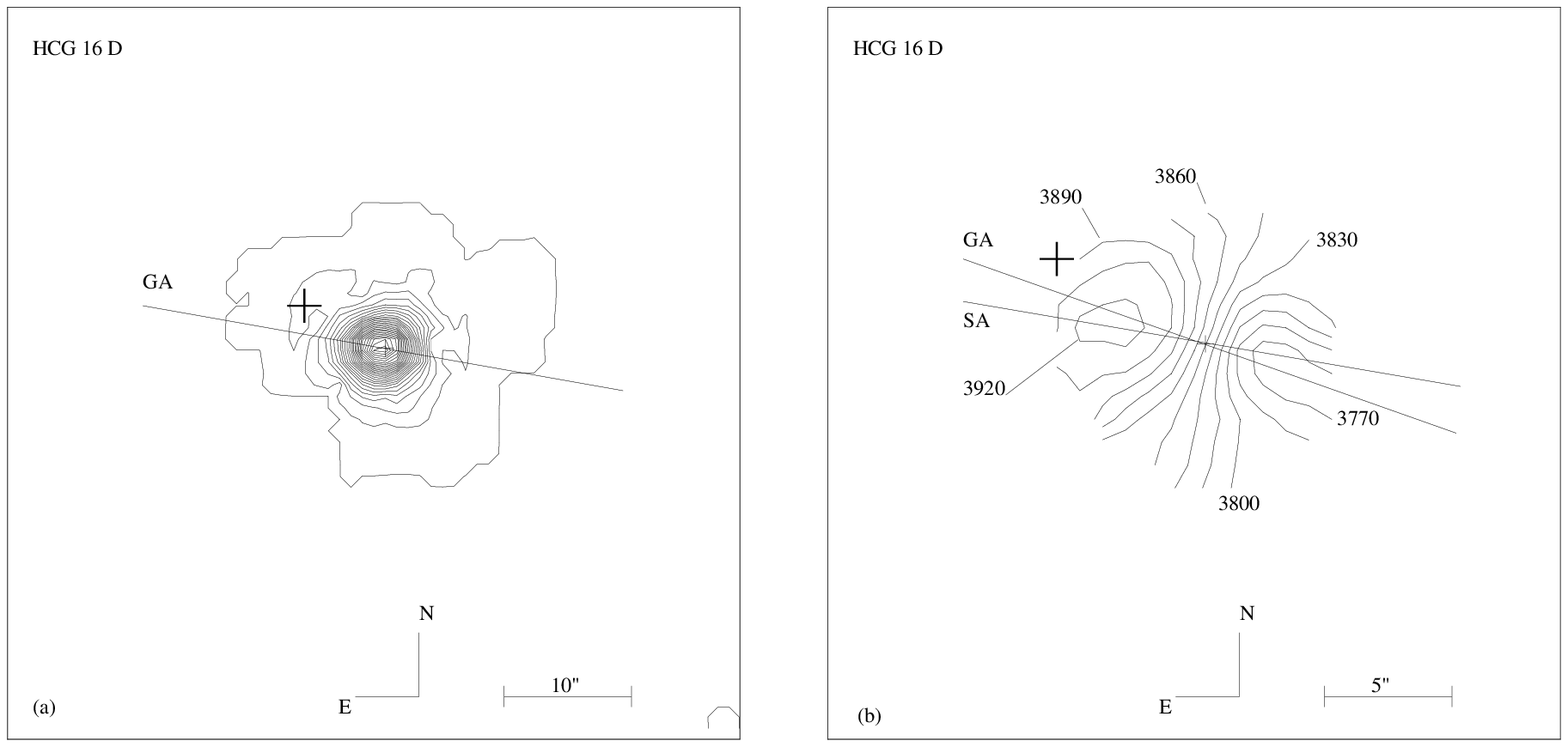]{(a)
The monochromatic iso-intensity map for H16d, calibrated in units of
10$^{-17} erg s^{-1} cm^{-2} arsec^{-2}$.  The lowest level is 0.7.
and the step is 10.  The small cross marks the monochromatic center of
the galaxy. The continuous line marks the gas major axis, measured from
the morphology of the monochromatic image (GA, which is almost
coincident with the stellar axis - see Table 2).  (b) The velocity
field built from the H$\alpha$-line emission. The small cross
represents the kinematic center.  The two continuous lines mark the
kinematic (GA) and stellar (SA) major axes.  In both pannels the large
cross, 7 $\arcsec$ to the east of the main center, indicates the
position of the second nucleus of the galaxy (see section 3.1).  
\label{fig11}}

%
%

\figcaption[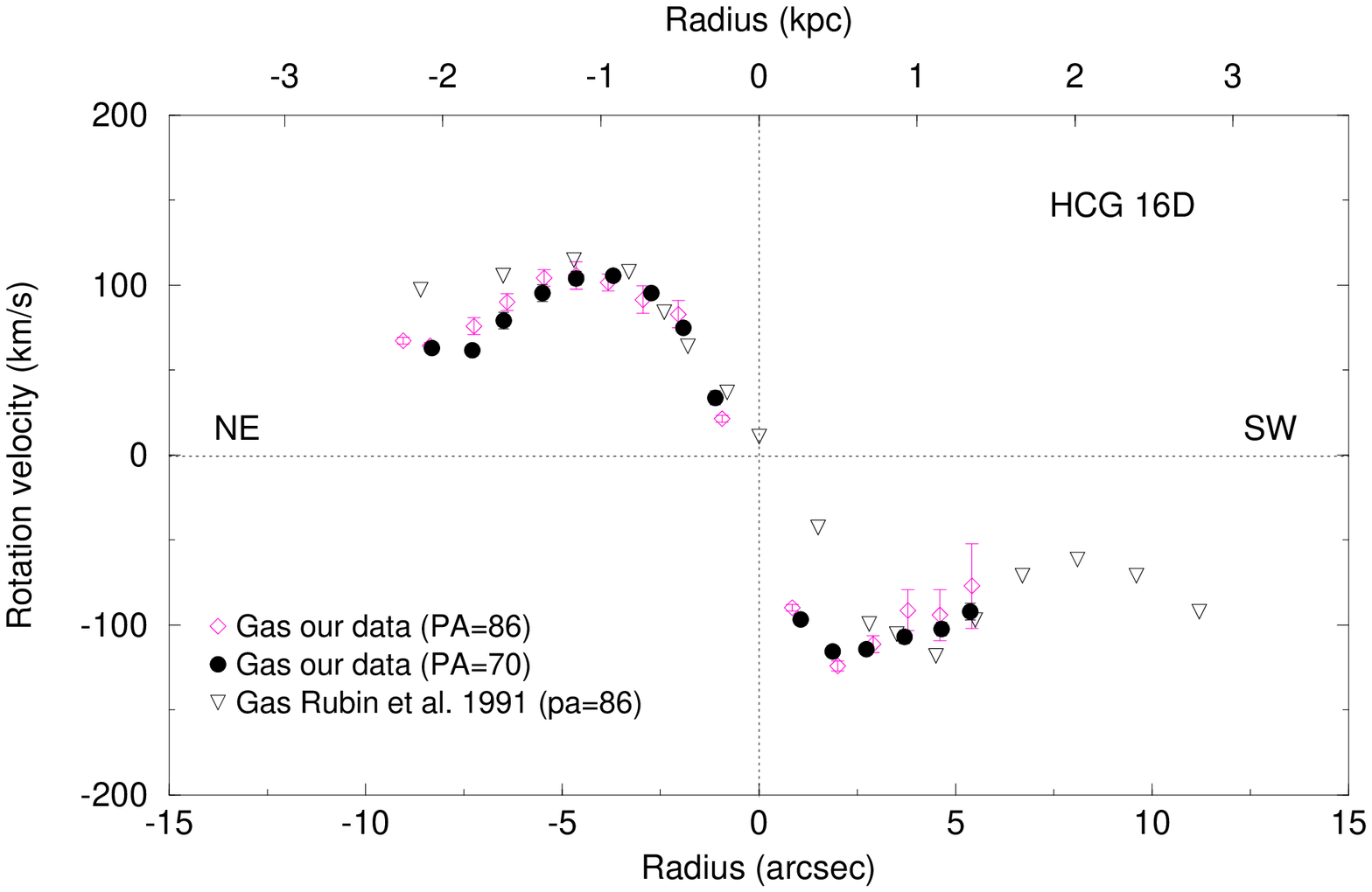]
{Rotation curve for H16d. De-projected velocities are plotted for two
values of major axis position angles, that determined from the velocity field
of HCG 16d (70\degr, solid dots) and the PA
obtained by RHF91 from the photometry
of the galaxy (86\degr, open diamonds). 
Overplotted as open triangles are the velocities 
measured by RHF91 along the photometric major axis of the galaxy.
\label{fig12} }

%
%
\figcaption[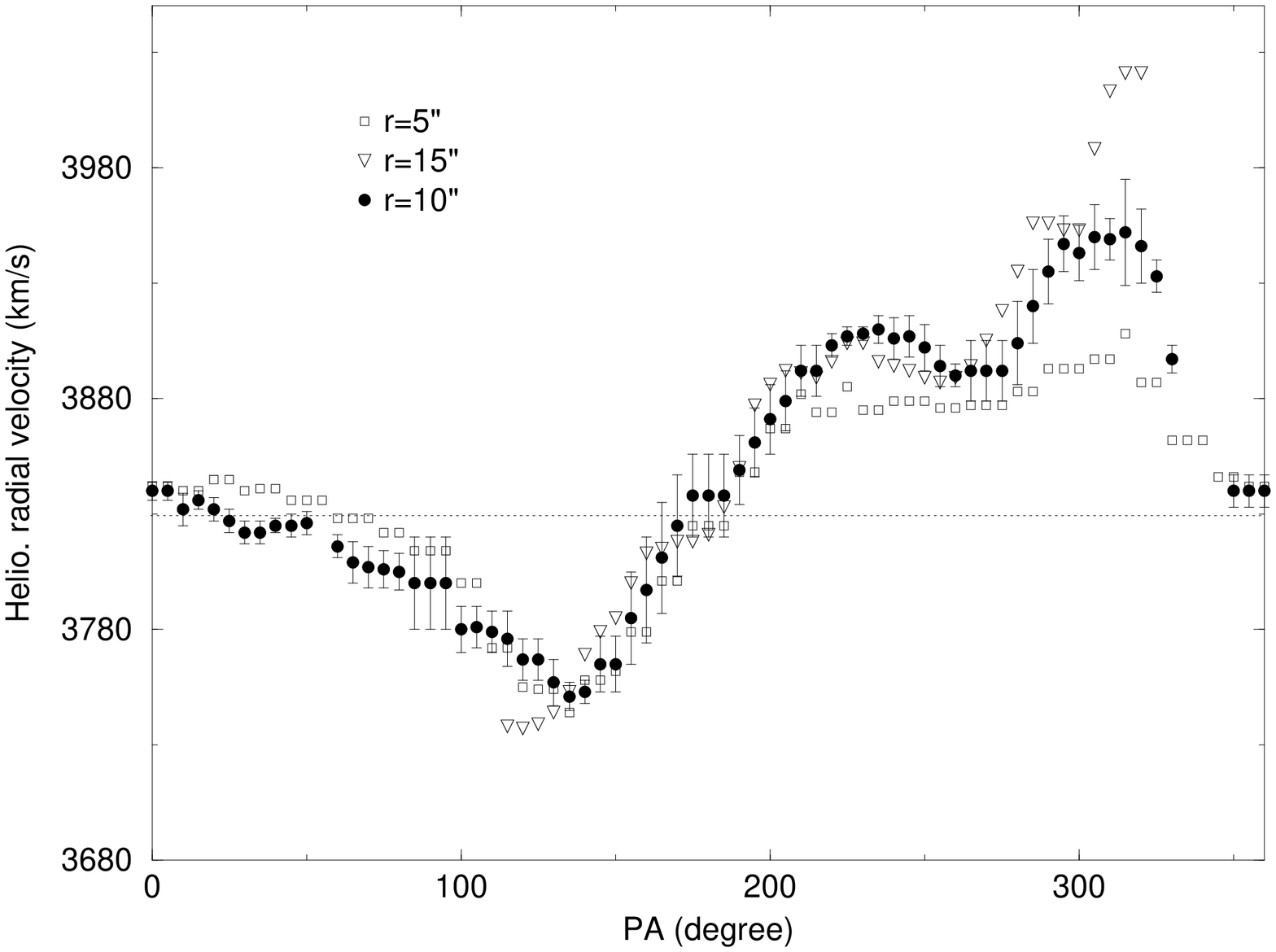]
{Heliocentric radial velocity {\it vs.} position angle for
H16c, for three different values of r, where r is the distance to
the center of the galaxy. The effect of a possible second
gas component is seen between PAs 220$\degr$ and 280$\degr$.\label{fig13}}

%


\plotone{fig1.ps}

\plottwo{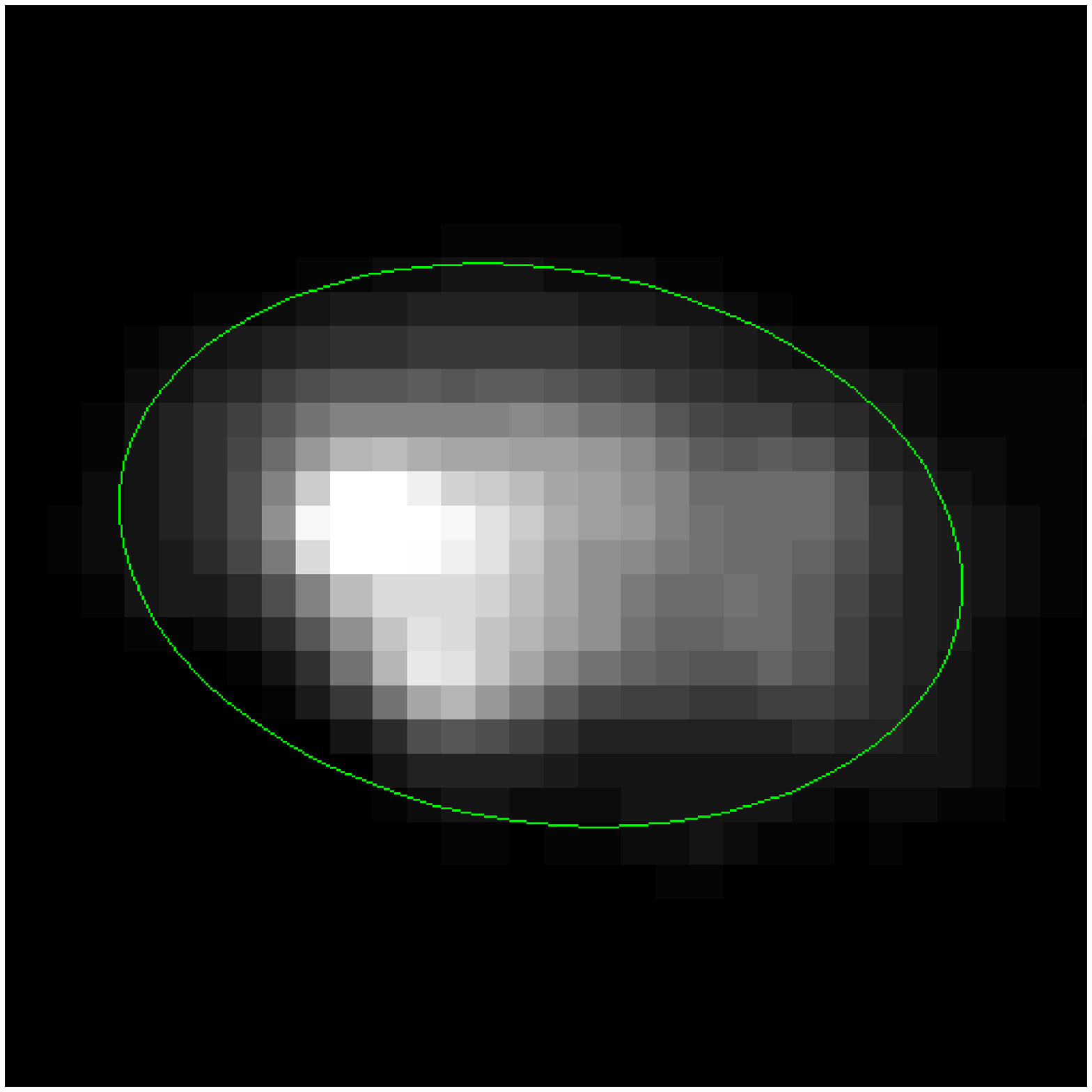}{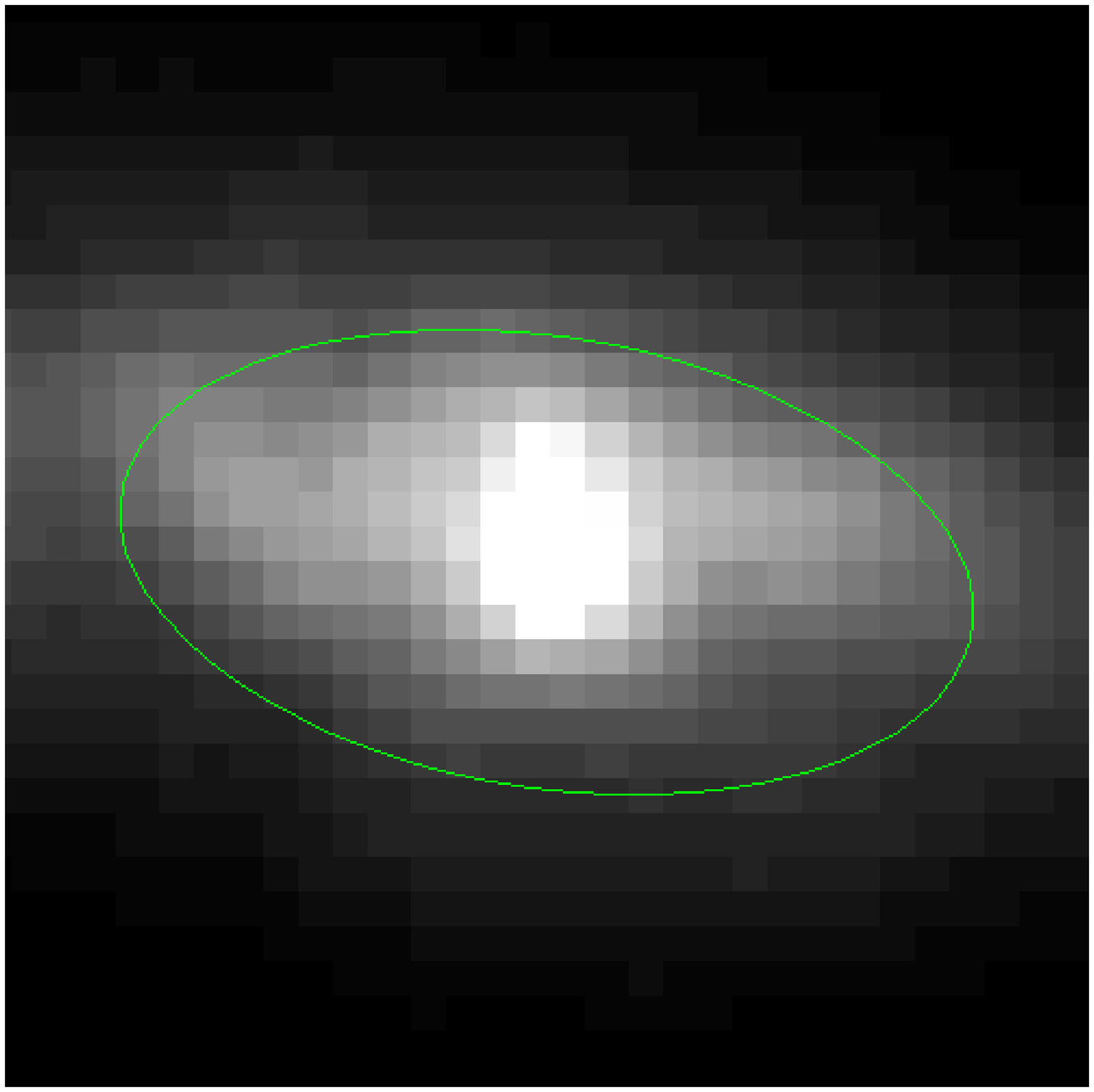}

\centerline{Fig. 2}

\plotone{fig3.ps}

\centerline{Fig. 3}

\plotone{fig4.ps}

\centerline{Fig. 4}

\plotone{fig5.eps}

\centerline{Fig. 5}

\plotone{fig6.eps}

\centerline{Fig. 6}

\plotone{fig7.eps}

\centerline{Fig. 7}

\plotone{fig8.eps}

\centerline{Fig. 8}

\plotone{fig9.ps}

\centerline{Fig. 9}

\plotone{fig10.eps}

\centerline{Fig. 10}

\plotone{fig11.ps}

\centerline{Fig. 11}

\plotone{fig12.eps}

\centerline{Fig. 12}

\plotone{fig13.eps}

\centerline{Fig. 13}

\end{document}